\providecommand{\tabularnewline}{\\}

\documentclass[referee,a4paper,12pt,structabstract,english]{swsc2}
\usepackage[T1]{fontenc}
\usepackage{amstext}
\usepackage{amssymb,amsmath}
\usepackage{graphicx,txfonts,subfigure,lineno,url,multirow,array,epsfig}
\usepackage[backref]{hyperref}
\usepackage[authoryear,round]{natbib}
\PassOptionsToPackage{normalem}{ulem}
\usepackage{ulem}

\hypersetup{colorlinks=true,citecolor=cyan,urlcolor=cyan,linkcolor=blue}

\title{Image patch analysis of sunspots and active regions. I. Intrinsic dimension and correlation analysis}
\titlerunning{Image patch analysis of active regions: Intrinsic dimension and correlation}
\authorrunning{Moon et al}
\author{Kevin R. Moon\inst{1,*} \and Jimmy J. Li\inst{1} \and V\'{e}ronique Delouille\inst{2} \and Ruben De Visscher\inst{2} \and Fraser Watson\inst{3} \and Alfred O. Hero III\inst{1}}
\institute{Electrical Engineering and Computer Science Department, University of Michigan 
\\ *Corresponding author: \texttt{krmoon@umich.edu} \and  SIDC, Royal Observatory of Belgium \and National Solar Observatory, Boulder, CO}
\usepackage{multirow}

\makeatother

\usepackage{babel}
\begin{document}
\abstract  
{The flare-productivity of an active region is observed to be related to its spatial complexity. Mount Wilson or McIntosh sunspot classifications measure such complexity but in a  categorical way, and may therefore not use all the information present in the observations. Moreover, such categorical schemes hinder a systematic study of an active region's evolution for example.    }
{
We propose fine-scale quantitative descriptors for an active region's complexity  and relate them to the Mount Wilson classification. We analyze the local correlation structure within continuum and magnetogram data, as well as the cross-correlation between continuum and magnetogram data.      
}         
{We compute the intrinsic dimension, partial correlation and canonical correlation analysis (CCA) of image patches of continuum and magnetogram active region images taken from the SOHO-MDI instrument.  We use masks of sunspots derived from continuum as well as larger masks of magnetic active regions derived from magnetogram to analyze separately the core part of an active region from its surrounding part.       
}        
{        
We find relationships between the complexity of an active region as measured by its Mount Wilson classification and the intrinsic dimension of its image patches. Partial correlation patterns exhibit approximately a third-order Markov structure. CCA reveals different patterns of correlation between continuum and magnetogram within the sunspots and in the region surrounding the sunspots.    
}        
{Intrinsic dimension has the potential to distinguish simple from complex active regions. These results also pave the way for patch-based dictionary learning  with a view towards automatic clustering of active regions.
}
\maketitle

\section{Introduction}

\label{sec:intro}Active regions (AR) in the solar atmosphere have
intense and intricate magnetic fields that emerge from subsurface
layers to form loops which extend into the corona. When active regions
undergo external forcing such as flux emergence and rearrangement,
the system may destabilize. The stored magnetic energy is then suddenly
released as accelerated particles (electrons, protons, ions) and an
increase in radiation called a \emph{flare} is observed across the
entire electromagnetic spectrum~\citep{1991VA.....34..353P}. 

The morphology of sunspots is correlated with flare occurrence and
has therefore received a lot of attention. The Mount Wilson classification
scheme~\citep{1919ApJ....49..153H} groups sunspots into four main
classes based on the magnetic structure, that is, on the relative locations and sizes of concentrations of opposite polarity magnetic flux. The sunspots with simplest morphology belong to the unipolar
$\alpha$ and the bipolar $\beta$ groups. More complex morphologies
are described as $\beta\gamma$ when a bipolar sunspot is such that
a single north-south polarity inversion line cannot divide the two
polarities. When a $\beta\gamma$ sunspot group contains in addition
a $\delta$ spot, that is, umbrae of different polarities inside a
single penumbra, it is labeled as a $\beta\gamma\delta$ group. The
presence of a $\delta$ configuration, where large values of opposite
polarity exist close together, was identified as a warning of the
build up of magnetic energy stress with an increased probability of
a large flare \citep{1985SoPh...96..293M,2000ApJ...540..583S}.
\cite{1990SoPh..125..251M} proposes another classification scheme containing 60 classes, thus describing the magnetic structure in greater details.  The McIntosh classification is the basis for several flare
forecasting methods which estimate the flare occurrence rate from historical
records of flares and active region classes \citep{Bornmann1994flares}, possibly
combining such information with observed waiting time distribution between flares \citep{2002SoPh..209..171G,2012ApJ...747L..41B}. 

The McIntosh and Mount Wilson classifications are in general carried
out manually, and this results in inconsistencies that stem from human
observation bias as well as non-reproducible catalogs. To overcome
these caveats, some supervised machine learning methods have been
proposed to automatically classify sunspot groups according to these
schemes. \cite{2013Stenning} extract various measurements from continuum
and magnetogram images, and then feed these into a machine learning
classifier which reproduces the Mount Wilson classification. \cite{2008SoPh..248..277C}
employ neural networks and supervised classification techniques to
reproduce the McIntosh scheme and use those results in a flare
forecasting system \citep{2009SpWea...7.6001C}. While these approaches
reduce the human bias, they do not use the information present in
sunspot images in an optimal way and make the study of AR dynamic
behavior impractical. 

Several attempts were made  to find quantitative descriptors of an active region's
complexity. \cite{2005ApJ...631..628M} showed that fractal dimension of an active region alone cannot distinguish between the various Mount Wilson classes. The generalization to multifractal spectrum, where each scale has its own fractal dimension, allowed to study in greater details the evolution of active region in view of distinguishing between quiet and flare-productive active regions. Box counting methods \citep{2005SoPh..228....5G,2005SoPh..228...29A,2008SoPh..248..297C} as well as more accurate methods based on continuous wavelet transform \citep{2010ApJ...717..995K,2010ApJ...722..577C} were employed. Continuous wavelet transforms and energy spectrum were also used with a similar purpose in \cite{2008SoPh..248..311H,2010AdSpR..45.1067M}. 

Wavelet basis functions act as a microscope to describe local discontinuities
and gradients in an image, and~\cite{Ireland2008mra} used two multiresolution analyses to compute at various length scales the gradients of the magnetic field along lines separating opposite polarities. Using a data set of about 10 000 magnetogram images, they showed that, at all length scales,  those gradients increase going from $\alpha$ to $\beta$, $\beta\gamma$, and $\beta\gamma\delta$ classes. 

\begin{figure}
\centering

\includegraphics[width=0.33\textwidth]{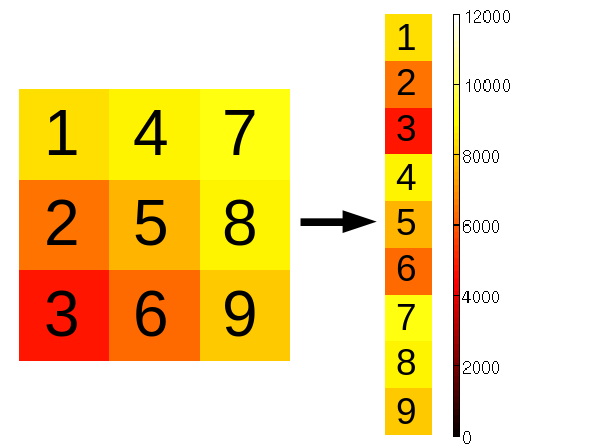}

\caption{An example patch from the edge of a sunspot in a continuum image and
its column representation.\label{fig:patch}}

\end{figure}

However, a wavelet analysis is known to generate artifacts
due to the particular shape of the specific wavelet functions. Signal representations based on a set of redundant functions called a \emph{dictionary}, were therefore introduced \citep{mallat_matching}. \cite{elad2006} proposed the use of a small sized dictionary to find a sparse representation of \emph{patches}. Specifically, a patch is a $m \times m$-pixel neighborhood, and a
patch analysis of a $n$-pixel image will process the $m^2 \times n$
data matrix that collects the overlapping patches. See Figure~\ref{fig:patch}
for a representation. 

As an example of image patch analysis, \cite{elad2006} considered the problem of denoising an image corrupted by additive Gaussian noise. They computed a sparse representation of patches over a dictionary, thus effectively denoising the patches. The dictionary itself may either be fixed \emph{a priori} or \emph{learned} from the corrupted patches. An estimate of the noise-free image is then obtained by averaging the denoised overlapping patches. \cite{elad2006} showed that dictionary learning methods based on patch analysis are more flexible and provide superior results in the context of image denoising.

In this paper, we carry out a patch analysis of a set of sunspots
and active region magnetogram images that span the four main Mount
Wilson classes. We estimate the intrinsic dimension of the local patches,
and show how it relates to the Mount Wilson classification. We also
study patterns of local correlation using partial correlation and
canonical correlation analysis, which reveal some characteristics
of simple and more complex active regions. Such analysis also serves
as a preparation to an unsupervised clustering of active region using  patch-based dictionary learning which will
be presented in a companion paper \citep{MoonPaper2}.

Section~\ref{sec:data} describes our data set. Unlike previous works, our approach combines
information from two modalities: photospheric continuum images and
magnetograms, both obtained by the \emph{Michelson Doppler Imager}
(MDI) on board the \emph{Solar and Heliospheric Observatory} (SOHO)
spacecraft. We consider 424 active regions spanning the four main
Mount Wilson classes. We use SMART masks~\citep{higgins2011solar}
to delineate the boundaries of magnetic active regions, and the
STARA algorithm \citep{2011masks} which provides masks for umbrae and penumbrae
from the continuum images. These two masks enable us to differentiate
between pixels belonging to the actual sunspots and pixels featuring
the region surrounding the sunspots.

In Section~\ref{sec:dimension}, the intrinsic dimension of the image
patches extracted from the two modalities is estimated using both
linear and non-linear methods. The linear method relies on Principal
Component Analysis (PCA) \citep{Jolliffe2002}, while the non-linear
method relies on a $k$-Nearest Neighbor graph approach \citep{costa2006determining,carter2010local}.
The latter method also estimates the local intrinsic dimension, which has several advantages over a global estimate. We show that the intrinsic dimension is related to the complexity
of the sunspot groups.

Section~\ref{sec:correlation} identifies the spatial and modal interactions
of the patches at different scales by estimating the partial correlation
and by using canonical correlation analysis (CCA) \citep{muller82,nimon:cca}.
This gives insight about relationships that may exist between active
region complexity and the correlation patterns.

This paper expands and refines some of the work in~\cite{moon2014icip}.
Whereas~\cite{moon2014icip} used fixed size square pixel regions
centered on the sunspot group as input to the analyses, in this paper
SMART detection masks are used. A larger set of images is considered
in all methods which enables us to analyze the relationships of intrinsic
dimension and correlation with AR complexity. We also explore the
partial correlation of patches which was not included in \cite{moon2014icip}.

\section{Data}

\label{sec:data}The data used in this study are taken from the \emph{Michelson Doppler Imager}
(MDI) instrument~\citep{1995MDI} on board the SOHO Spacecraft. 

Within the time range of 1996-2010, we select a set of 424 ARs as follows. Using the information from the Solar Region
Summary reports compiled by the Space Weather Prediction Center of
NOAA~\url{http://www.swpc.noaa.gov/ftpdir/forecasts/SRS/}, we consider
ARs located within $30^\circ$ of the solar meridian. We looked at
a maximum of two-hundred instances per Mount Wilson types $\alpha, \beta, \beta\gamma,$
and $\beta\gamma\delta$. Out of this first selection, we removed
AR with a longitudinal extent smaller than four degrees, and finally we checked
if MDI continuum and magnetogram data were available. This provides
us with a number of ARs in each Mount Wilson class as given by Table~\ref{tab:mwilson}.
In our analysis, we also divide the ARs into two groups: simple ARs
($\alpha$ and $\beta$) and complex ARs ($\beta\gamma$ and $\beta\gamma\delta$). 

\begin{table}
\caption{Number of each AR per Mt. Wilson class. Simple ARs include $\alpha$
and $\beta$ groups while complex ARs are $\beta\gamma$ and $\beta\gamma\delta$
groups.\label{tab:mwilson}}
\centering

\begin{tabular}{|c|c|c|c|c|c|c|c|}
\cline{2-8} 
\multicolumn{1}{c|}{} & $\alpha$ & $\beta$ & $\beta\gamma$ & $\beta\gamma\delta$ & Simple & Complex & Total\tabularnewline
\hline 
Number of AR & 50 & 192 & 130 & 52 & 242 & 182 & 424\tabularnewline
\hline 
\end{tabular}

\end{table}

AR are observed using two modalities: photospheric continuum images
and magnetogram. SOHO-MDI provides almost continuous observations
of the Sun in the white-light continuum, in the vicinity of the Ni~{\sc i}
676.78~nm photospheric absorption line. These photospheric intensity
images are primarily used for sunspot observations. MDI data are available
in several processed ``levels''. We use level-1.8 images, and rotate
them with North up. SOHO provides two to four MDI photospheric intensity
images per day with continuous coverage since 1995. We also use the level-1.8 line-of-sight (LOS) MDI magnetograms, recorded with a nominal cadence of 96 minutes. The magnetograms show the magnetic
fields of the solar photosphere, with negative (represented as black)
and positive (as white) areas indicating opposite LOS magnetic-field
orientations.

As stated in Section~\ref{sec:intro}, SMART masks~\citep{higgins2011solar}
are used to determine the boundaries of magnetic active regions from
MDI magnetograms. Those masks are applied also on continuum images
to determine the surrounding part of the sunspot that is affected
by magnetic fragments as seen in magnetogram images. Similarly, the
STARA algorithm~\citep{2011masks} provides masks for sunspots (umbrae
and penumbrae) from MDI continuum and those masks are applied on magnetogram
images to determine the AR cores corresponding to the sunspots. Combining
these two types of masks provides thus two sets of pixels within each
AR: those belonging to the \emph{sunspots} themselves as found by
STARA and those belonging to the \emph{magnetic fragments} (or background)
within an AR as found by the difference set between the SMART and
STARA masks. 

As in~\cite{moon2014icip} we use image patch features to account
for spatial dependencies using square patches of pixels. Thus if a
SMART mask of an image has $n$ pixels and we use a $m\times m$ patch,
the corresponding continuum data matrix $\mathbf{X}$ is $m^{2}\times n$ where
the $i$th column contains the pixels in the patch centered at the
$i$th pixel. The magnetogram data matrix $\mathbf{Y}$ is formed in the same
way and the full data matrix is $\mathbf{Z}=\left(\begin{array}{c}
\mathbf{X} \\ \mathbf{Y}\end{array}\right)$ with size $2m^{2}\times n$. We let $\mathbf{z}_{i}$ denote the $i$th column
of $\mathbf{Z}$. The images from both modalities are also normalized prior to analyzing them. 

In image patch analysis, the size of the patch should be no larger than the smallest feature that is to be captured. Otherwise, the relevant feature may be suppressed. Additionally, large patches lead to high-dimensional estimates which suffer in accuracy from "the curse of dimensionality," which refers to the fact that the number of observations must increase at least linearly in the number of parameters for accurate estimates to be possible in statistical inference \citep{buhlmann2011statistics}. Since some sunspot and active region features can be quite
small and to limit the effects of high dimensionality on our analysis, we primarily use $3\times3$ patches in each modality although
larger patches are used in Section~\ref{sec:correlation} when analyzing
spatial correlations in the images.

\section{Intrinsic Dimension Estimation}

\label{sec:dimension}The goal of this section is to determine the
number of intrinsic parameters or degrees of freedom required to describe
the spatial and modal dependencies using image patches. We consider
$3\times3$ patches within both the continuum and magnetogram images
giving an extrinsic dimension of 18. The intrinsic dimension will
determine how redundant these 18 dimensions are. In addition, intrinsic
dimension provides an indicator of complexity which we compare against
the Mount Wilson classification, similarly to what~\cite{2005ApJ...631..628M} and \cite{Ireland2008mra}
did using fractal dimension and gradient strength along polarity separating lines, respectively. More details on the concept of intrinsic
dimension on manifolds are included in Appendix~\ref{sub:knn}. 

It is also important to know whether \emph{linear }analyses can be
accurately applied to the data or whether \emph{non-linear} techniques
are required. Linear methods have been applied successfully to solar images before such as in \cite{2013SoPh..283...31D}. However, it is not guaranteed that natural images are best represented using linear methods as there are cases where non-linear models have superior performance \citep{dobigeon2014nonlinear}. Thus this is important to investigate both for further analysis of the data as in \cite{MoonPaper2} 
and for the correlation analysis in Section~\ref{sec:correlation}. If the 
data lie on a nonlinear subspace and we perform a linear analysis of the data 
(e.g. partial correlation, canonical correlation analysis, or principal component analysis),
 then the results will be only a linear approximation of the true relationships and 
dependencies of the data. Nonlinear methods of analysis would be necessary to obtain higher 
accuracy in this case. To answer this question, we estimate the local intrinsic
dimension using a method appropriate for linear subspaces and a method
appropriate for any (linear or non-linear) smooth subspace and then
compare the results.

\subsection{PCA: A Linear Estimator}

Principal Component Analysis (PCA) \citep{Jolliffe2002} finds a set
of linearly uncorrelated vectors (principal components) that can be
used to represent the data. PCA has been used previously for various
purposes in solar-physics and space-weather literature, e.g.
to study the background and sunspot magnetic fields \citep{2004SoPh..225....1L,2008SoPh..248..247C,2012MNRAS.424.2943Z},
for analysis of solar wind data \citep{2014JGRA..119.4544H}, or to
reduce dimensionality \citep{2007A&A...466..347D}. 

In PCA, the principal components are the eigenvectors of the covariance
matrix $\Sigma$:
\[
\Sigma=\left(\begin{array}{cc}
\Sigma_{\mathbf{xx}} & \Sigma_{\mathbf{xy}}\\
\Sigma_{\mathbf{yx}} & \Sigma_{\mathbf{yy}}
\end{array}\right),
\]
where $\mathbf{x}$ and $\mathbf{y}$ are random vectors of dimension
$9$, $\mathbf{x}$ being a patch from the continuum image, and $\mathbf{y}$
the corresponding patch from the magnetogram image. The eigenvalues
indicate the amount of variance accounted for by the corresponding
principal component. A linear estimate of intrinsic dimension is the
number of principal components that are required to explain a certain
percentage of the variance. 

By nature, PCA is a global operation and so it provides a global estimate
of the intrinsic dimension. We can obtain more local estimates by
performing PCA separately on the areas within the sunspots and on
the magnetic fragments. These areas are separated using the STARA
and SMART masks.

\subsection{$k$-NN: A General Estimator}

The general method we use is a $k$-Nearest-Neighbor ($k$-NN) graph
approach with neighborhood smoothing~\citep{costa2006determining,carter2010local}.
The intuition behind the method is that we grow the $k$-NN graph
from a point $\mathbf{z}_{i}$ by adding an edge from $\mathbf{z}_{i}$ to $\mathbf{z}_{j}$
if $\mathbf{z}_{j}$ is within the $k$ nearest neighbors of $\mathbf{z}_{i}$. The
growth rate of the total edge length of the graph is related to the
intrinsic dimension of the data in a way that enables us to estimate
it.

One advantage of the $k$-NN method, in contrast to global methods such as \cite{levina2004maximum},  is it provides an estimate of the
\emph{local }intrinsic dimension by limiting the growth of the graph
to a smaller neighborhood. This provides an estimate of intrinsic
dimension at each pixel location in the image which allows us to
more easily visualize the intrinsic dimension estimates. Additionally, when the number of samples within a region of interest is small (such as within a small sunspot), this local method provides more accurate estimates of intrinsic dimension than applying a global method (such as PCA) since the inclusion of the neighboring pixels results in a higher number of samples. Technical
explanation of the $k$-NN method and more details on local intrinsic
dimension are given in Appendix~\ref{sub:knn}.

\subsection{General Results}

\label{sub:dimresults}
\begin{figure}
\centering

\includegraphics[width=1\textwidth]{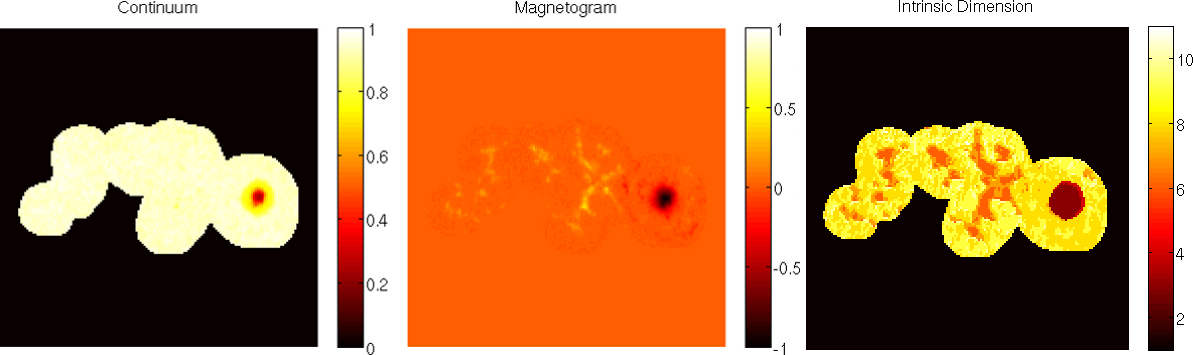}

\includegraphics[width=1\textwidth]{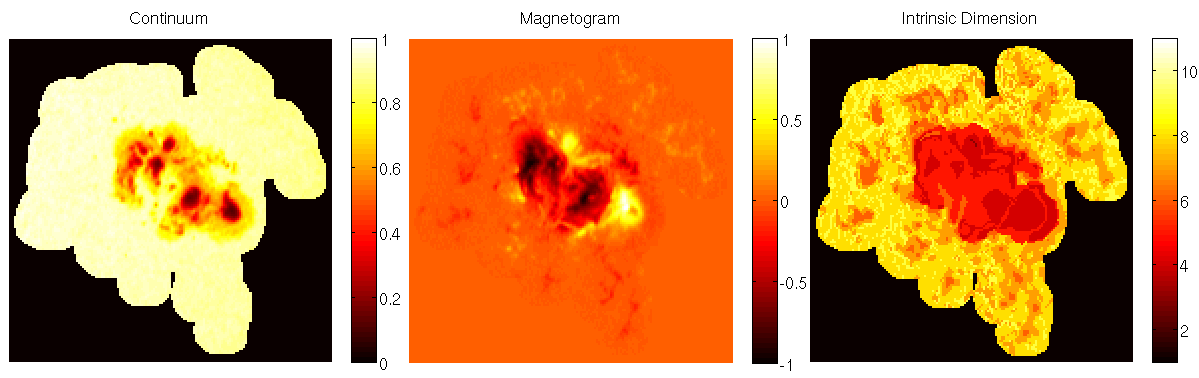}

\caption{Examples of the estimated local intrinsic dimension using the $k$-NN
method for an $\alpha$ group (top) and a $\beta\gamma\delta$ group
(bottom). Regions with more spatial structure have lower intrinsic
dimension.\label{fig:examples}}

\end{figure}

We estimate
the intrinsic dimension of the image patches within the sunspots and magnetic fragments
for all 424 ARs using both the $k$-NN approach and PCA, where the extrinsic dimension of the joint patches is 18. Figure~\ref{fig:examples}
shows two examples of the estimated local intrinsic dimension using
the $k$-NN method and the corresponding continuum and magnetogram
images. One set of images corresponds to an $\alpha$ group while
the other set is a $\beta\gamma\delta$ group. In these examples,
areas with more spatial structure, such as within the sunspots,
have lower intrinsic dimension. Fewer parameters are required to accurately
represent structured data than noise and so the intrinsic dimension
is lower. 

Table~\ref{tab:dimresults} provides the mean and standard deviation
of the intrinsic dimension estimates within the sunspots and magnetic
fragments. These statistics are also provided for ARs within the main
Mount Wilson classes ($\alpha$, $\beta$, $\beta\gamma$, and $\beta\gamma\delta$).
We provide PCA results for the cases where we estimate the intrinsic
dimension as the number of components required to explain $97\%$
and $98\%$ of the variance, respectively. For the $k$-NN method,
we provide the results in two ways. For one, we take the mean of local intrinsic dimensions within each image (separating the \lq sunspot' from the \lq magnetic fragments') and then calculate the mean and standard
deviation of these \emph{means}. The statistics in this category correspond
to the mean and standard deviation of the average intrinsic dimension
of each image and are more directly comparable to the PCA results.
However these results may be affected slightly by small sunspot groups.
For the other approach, we \emph{pool} all of the local estimates (again separating sunspots from magnetic fragments) 
and then calculate the mean and standard deviation. These results correspond
to the mean and standard deviation of the pixels within each region
and category and are less affected by small sunspot groups.

\begin{table}

\caption{Estimated intrinsic dimension results for different groups of ARs
in the form of mean$\pm$standard deviation. The complex ARs have
higher intrinsic dimension within the sunspots than the simple ARs
but lower intrinsic dimension within the magnetic fragments.\label{tab:dimresults}}
\centering

\begin{tabular}{|l|c|c|c|c|c|}
\cline{2-6} 
\multicolumn{1}{l|}{} & $\alpha$ & $\beta$ & $\beta\gamma$ & $\beta\gamma\delta$ & All\tabularnewline
\hline 
Sunspots $k$-NN, pooled & $3.9\pm1.2$ & $4.4\pm1.0$ & $4.4\pm0.9$ & $4.5\pm0.7$ & $4.3\pm1.0$\tabularnewline
\hline 
Sunspots $k$-NN, means & $4.0\pm1.0$ & $4.8\pm1.2$ & $4.4\pm0.6$ & $4.4\pm0.4$ & $4.5\pm1.0$\tabularnewline
\hline 
Sunspots PCA $97\%$ & $3.7\pm0.8$ & $4.4\pm0.9$ & $4.3\pm0.6$ & $4.2\pm0.5$ & $4.3\pm0.8$\tabularnewline
\hline 
Sunspots PCA $98\%$ & $4.5\pm0.9$ & $5.2\pm1.0$ & $5.1\pm0.8$ & $5.0\pm0.5$ & $5.0\pm0.9$\tabularnewline
\hline 
Fragments $k$-NN, pooled & $8.0\pm1.1$ & $8.0\pm1.2$ & $7.7\pm1.2$ & $7.6\pm1.2$ & $8.0\pm1.2$\tabularnewline
\hline 
Fragments $k$-NN, means & $8.0\pm0.4$ & $7.9\pm0.5$ & $7.6\pm0.5$ & $7.6\pm0.5$ & $7.8\pm0.5$\tabularnewline
\hline 
Fragments PCA $97\%$ & $7.7\pm1.6$ & $7.1\pm1.2$ & $6.2\pm1.2$ & $6.2\pm1.3$ & $6.8\pm1.4$\tabularnewline
\hline 
Fragments PCA $98\%$ & $9.1\pm1.6$ & $8.5\pm1.3$ & $7.5\pm1.2$ & $7.4\pm1.3$ & $8.1\pm1.4$\tabularnewline
\hline 
\end{tabular}

\end{table}

From Table~\ref{tab:dimresults}, it is clear that the intrinsic
dimension is lower within the sunspots than in the magnetic fragments
for all methods. This is expected as there is more spatial structure
within the images inside the sunspots than in the magnetic fragments,
especially in the continuum image. 

The average PCA estimate with a 97\% threshold and the average mean
$k$-NN estimate give similar results inside the sunspot while the
average 98\% PCA estimate is closest to the average mean $k$-NN estimate
within the magnetic fragments. If linear methods were not sufficient to represent the spatial and modal dependencies, we would expect the PCA results to be much higher than the $k$-NN results when using comparable thresholds as more linear than nonlinear components would be required to accurately represent the data. However, this close agreement between the general
and linear results suggests that linear methods are sufficient and that linear dictionary
methods would be appropriate for these data.

\subsection{Patterns Within the Mount Wilson Groups}
For both the PCA and $k$-NN methods, the average estimated intrinsic
dimension is lower within the sunspots in $\alpha$ groups than in
the more complex groups such as $\beta\gamma\delta$. This is consistent
with Figure~\ref{fig:examples} and may be related to the lower complexity
of $\alpha$ groups. These exhibit more spatially coherent images,
which can be described using a lower number of basis elements, and
hence have a lower intrinsic dimension. 

\begin{figure}
\centering

\includegraphics[width=0.5\textwidth]{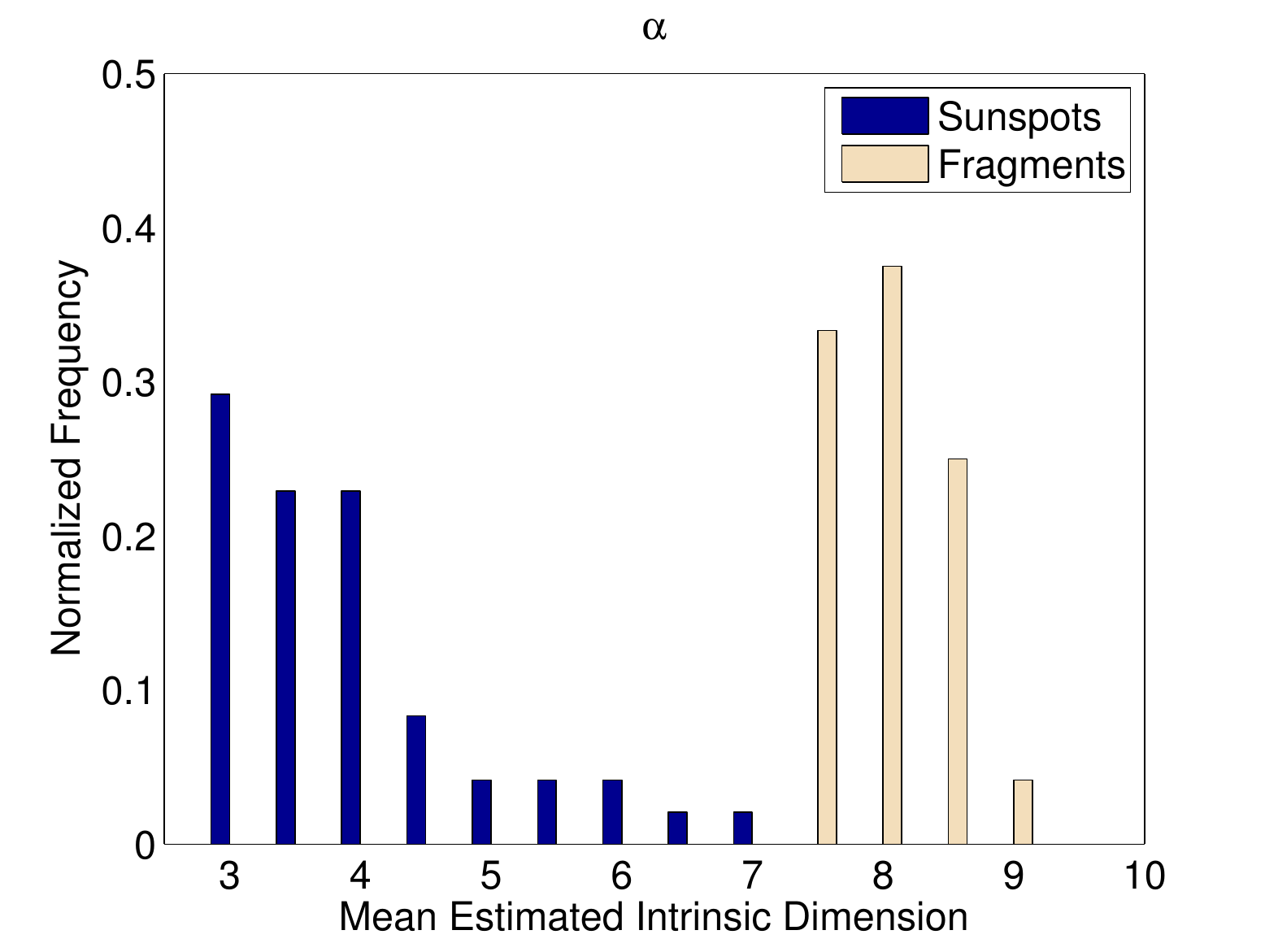}\includegraphics[width=0.5\textwidth]{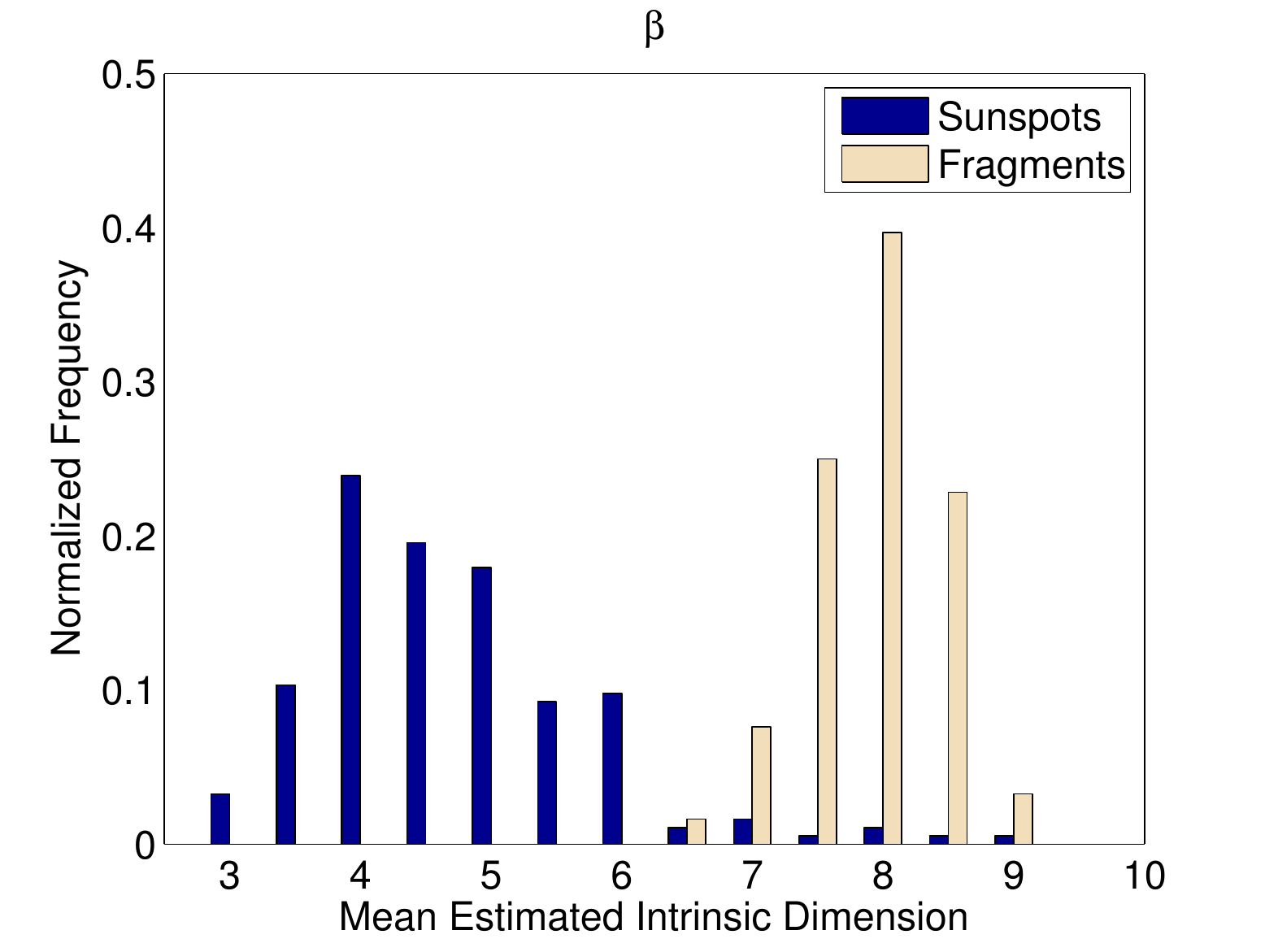}

\includegraphics[width=0.5\textwidth]{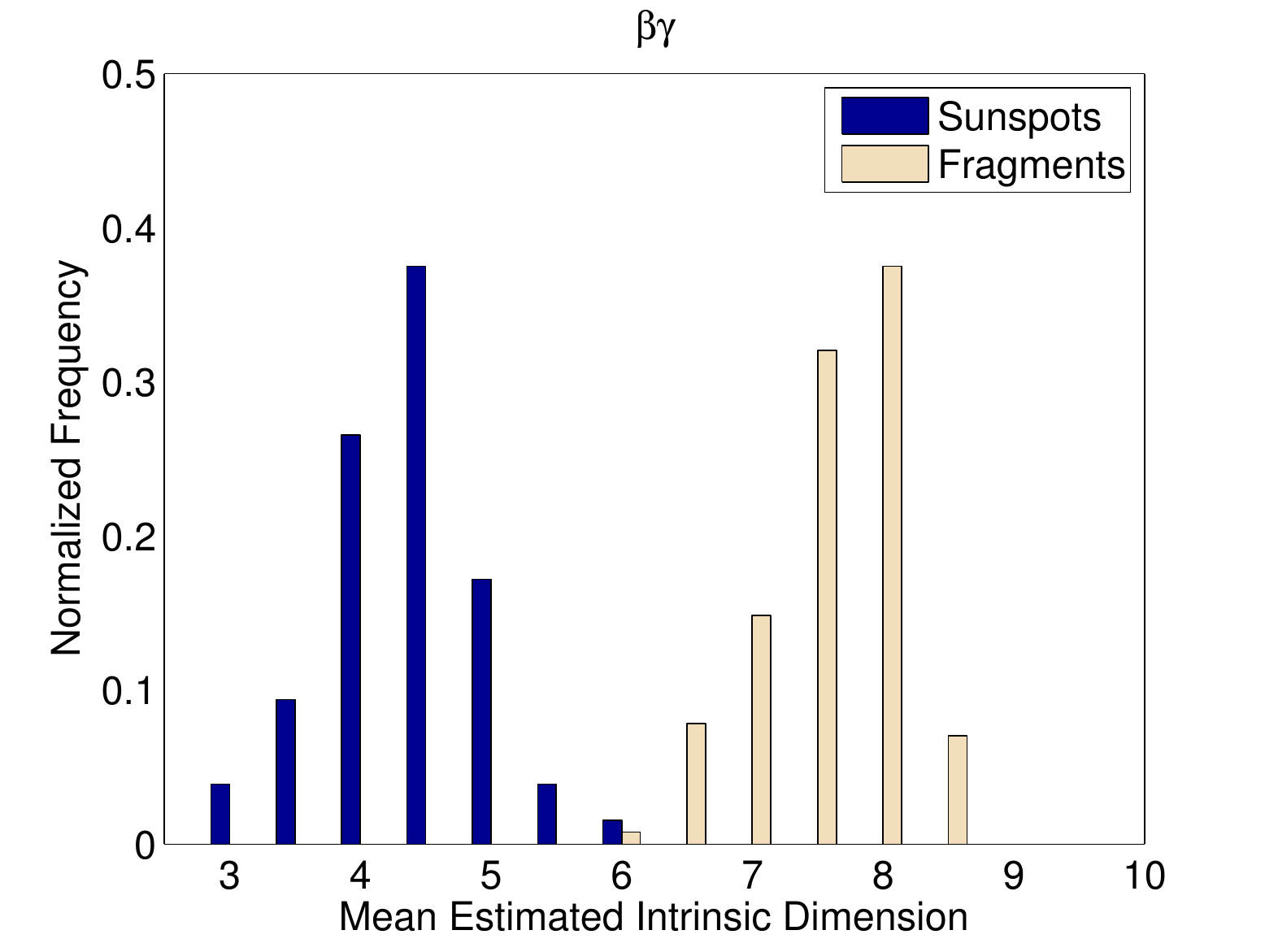}\includegraphics[width=0.5\textwidth]{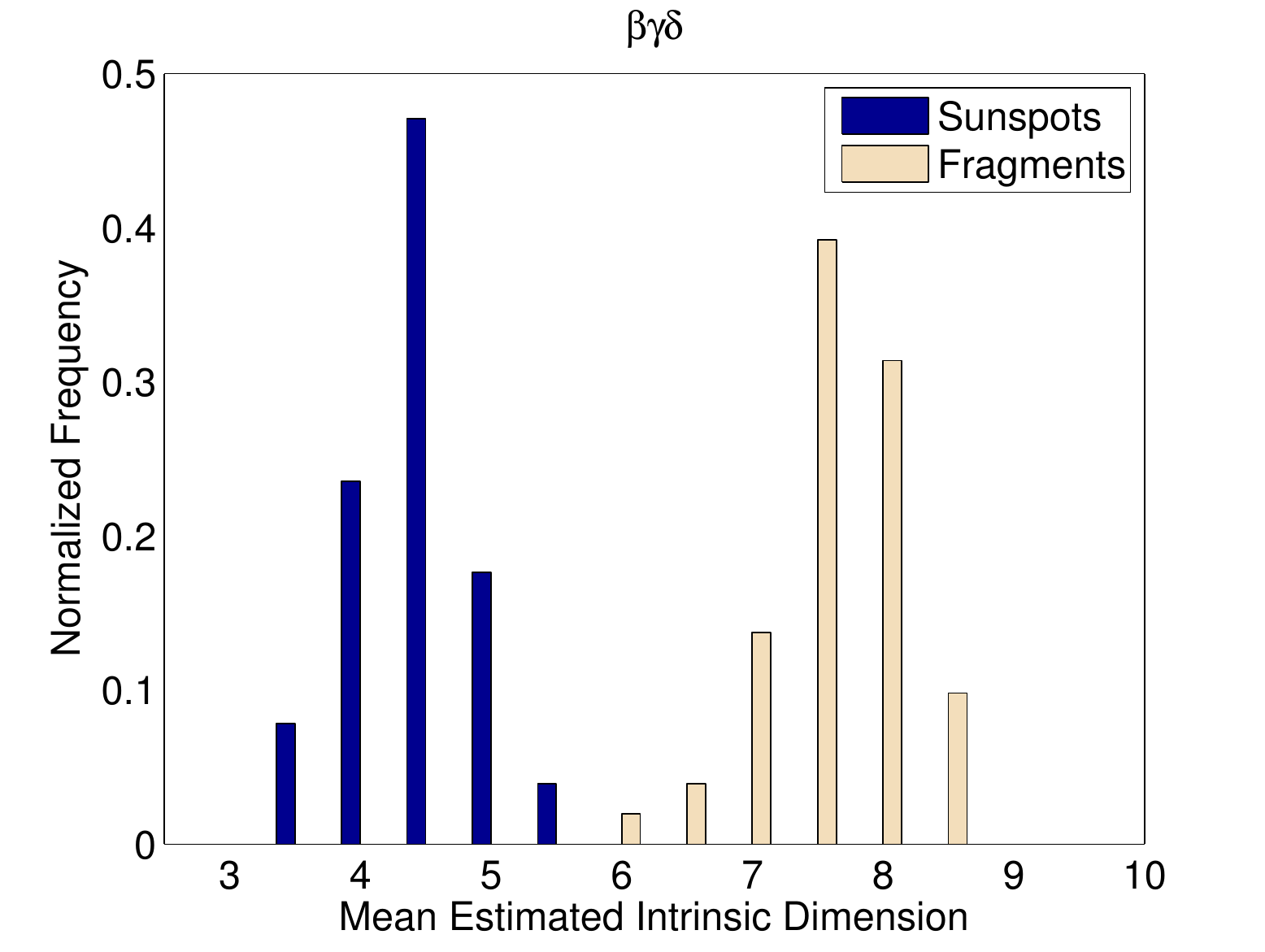}

\caption{Normalized histograms of mean estimated intrinsic dimension of $\alpha$,
$\beta,$ $\beta\gamma$, and $\beta\gamma\delta$ groups using the
$k$-NN method. The distributions of intrinsic dimension differ by
complexity with simpler AR groups having higher (resp. lower) intrinsic
dimension within the sunspot (resp. magnetic fragments). \label{fig:dimhist}
}

\end{figure}

\begin{figure}
\centering

\includegraphics[width=0.5\textwidth]{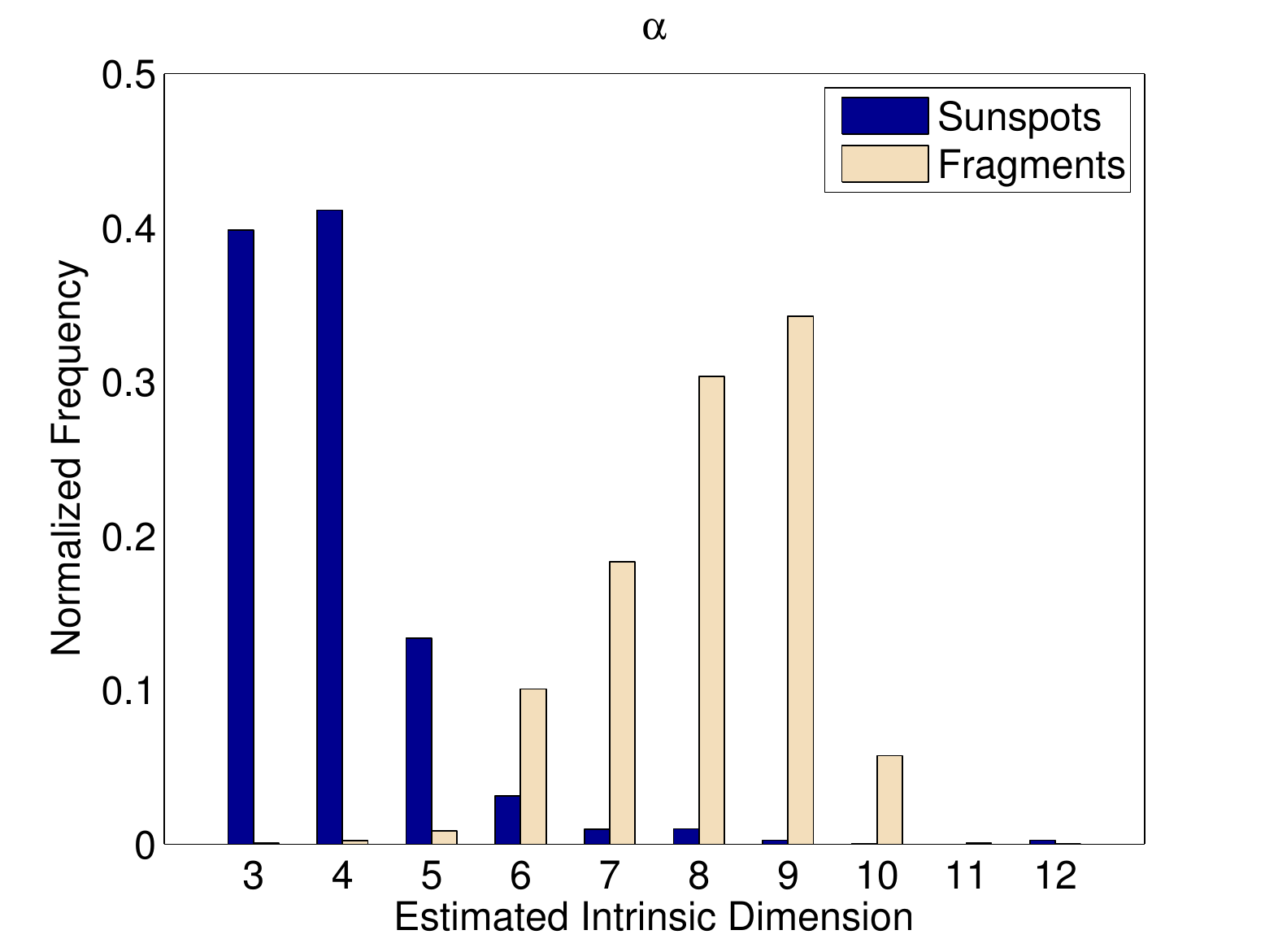}\includegraphics[width=0.5\textwidth]{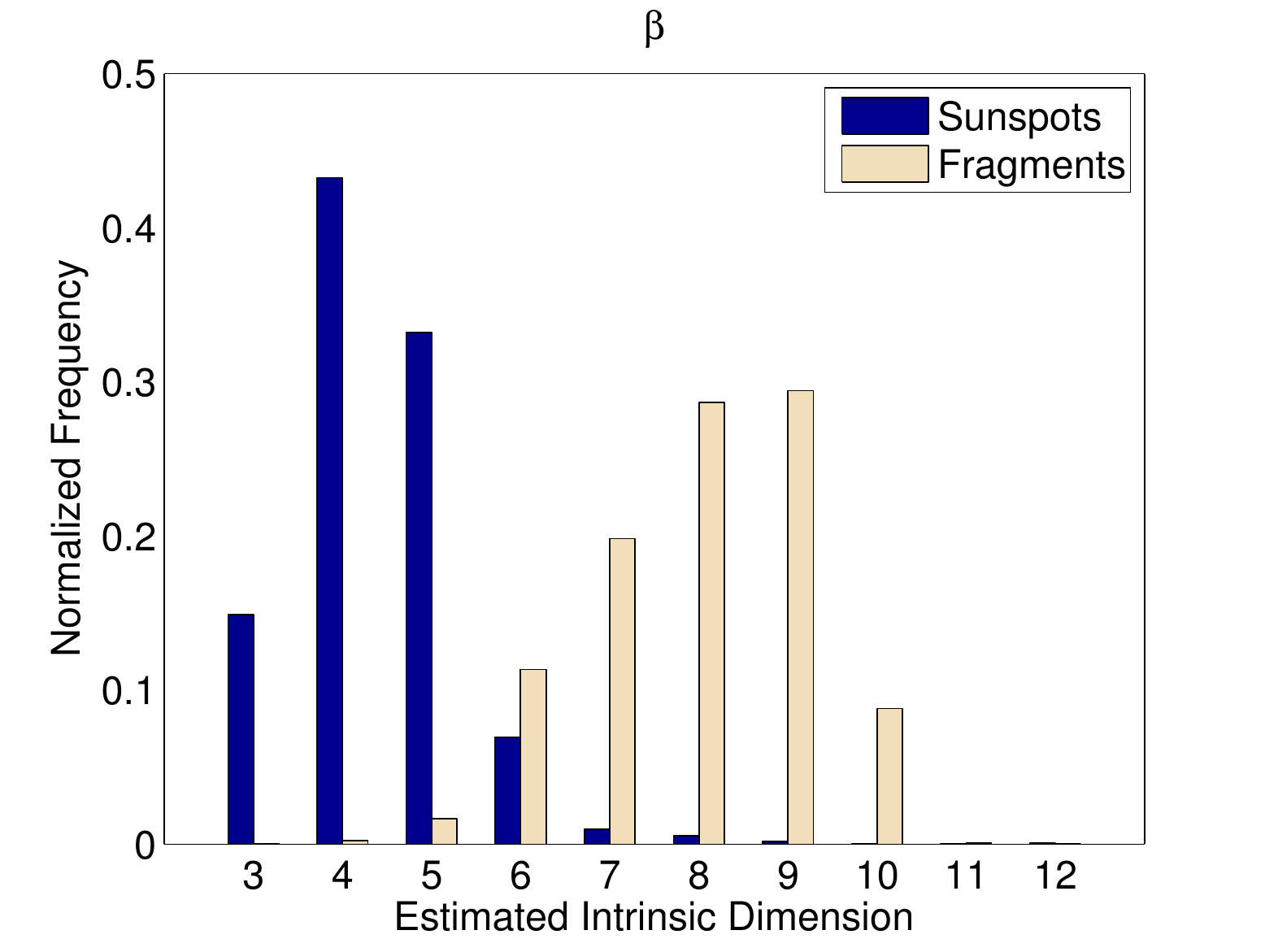}

\includegraphics[width=0.5\textwidth]{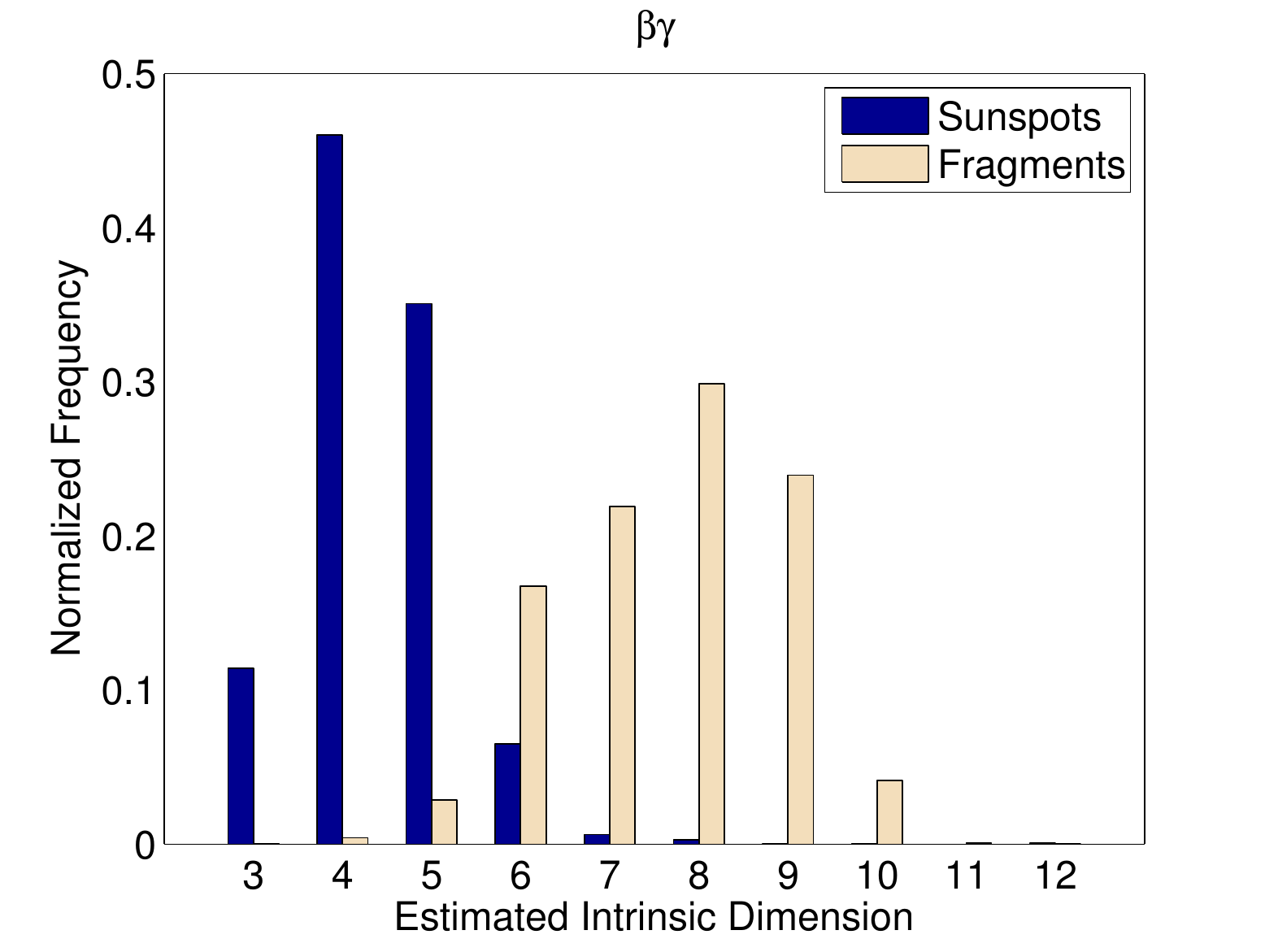}\includegraphics[width=0.5\textwidth]{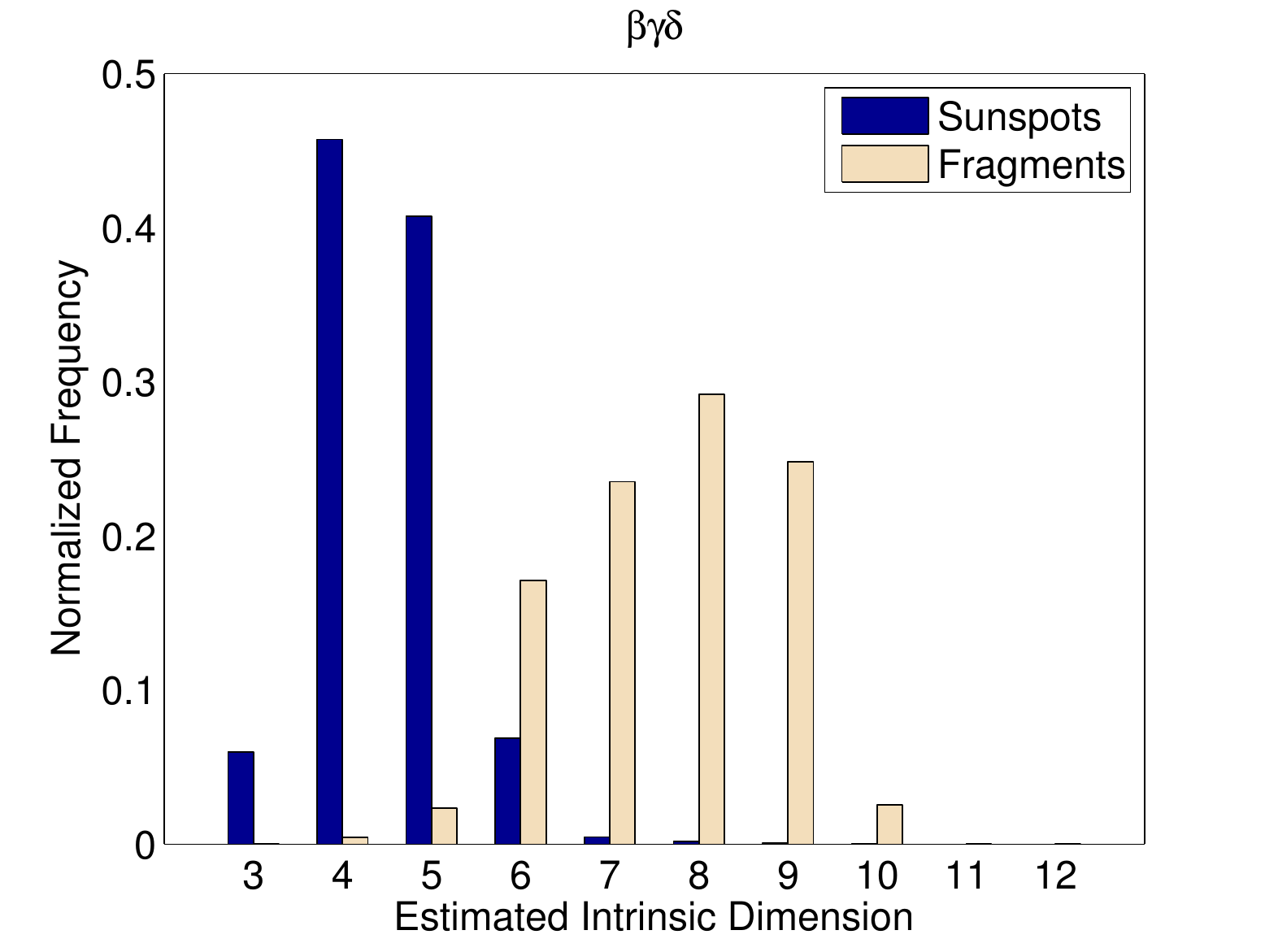}

\caption{Normalized histograms of pooled local estimates of intrinsic dimension
of $\alpha$, $\beta,$ $\beta\gamma$, and $\beta\gamma\delta$ groups
using the $k$-NN method. The distributions of intrinsic dimension
differ by complexity with simpler AR groups having higher (resp. lower)
intrinsic dimension within the sunspot (resp. magnetic fragments).
\label{fig:dimhist_pooled}}
\end{figure}

Within the magnetic fragments, the opposite trend occurs where the
less complex groups have higher intrinsic dimension. This suggests that the magnetic fragments are fewer, weaker, and less structured outside of the $\alpha$ and $\beta$ groups compared
to the more complex regions, leading  to a more noise-like background in their magnetic fragments. 
This hypothesis is supported
by the normalized histograms of the mean $k$-NN estimates of intrinsic
dimension and the normalized histograms of the pooled $k$-NN estimates
in Figures~\ref{fig:dimhist} and~\ref{fig:dimhist_pooled}. The histograms
of mean intrinsic dimension show that within the magnetic fragments, $\alpha$
groups generally have higher mean intrinsic dimension than $\beta\gamma\delta$
groups. In fact, no $\alpha$ groups have a mean intrinsic
dimension less than 7.5 within the magnetic fragments. However, the normalized
histograms of the individual patch estimates show a significant number
of patches with intrinsic dimension less than 7.5 within the fragments.
This suggests that for each $\alpha$ group, the majority of the patches
have higher intrinsic dimension in the magnetic fragments and are
thus more noise-like. In contrast, there are some $\beta\gamma\delta$
groups where the mean intrinsic dimension of the magnetic fragments
is lower (less than 7.5) and so these magnetic fragments are dominated
by patches with more structure. 

Table~\ref{tab:dimresults} also shows that the standard deviation
of the estimates within the sunspots decreases as the complexity increases
as measured by the Mount Wilson classification scheme. The histograms
in Figures~\ref{fig:dimhist} and~\ref{fig:dimhist_pooled} can be
used to determine the cause. From the histograms, it is clear that
within the sunspots the intrinsic dimension of $\alpha$ groups does
not have a Gaussian distribution. In this case, most of the estimates
are between 3 and 5. However, there are a significant number of outliers
with intrinsic dimension greater than 5. The presence of these outliers
contributes to the high standard deviation. This is in contrast to
the intrinsic dimension of $\beta\gamma$ and $\beta\gamma\delta$ groups inside the
sunspot which have fewer outliers and thus smaller standard deviations. 

The outliers in the $\alpha$ groups correspond to small sunspots.
The number of pixels within the $\alpha$ sunspots with average intrinsic
dimension $\geq6$ range between 10 and 53 with a median of 16. In
these cases, the spatial structure of the sunspots may be more similar
to the magnetic fragments than the spatial structure of larger sunspots.
Thus the intrinsic dimension is higher in the small sunspots. 

A similar phenomenon occurs within the $\beta$ groups. Note that
in Table~\ref{tab:dimresults}, the average and standard deviation
of the mean intrinsic dimension of the $\beta$ groups within the
sunspots is higher than for all other groups. This is also caused
by a few outliers that have high average intrinsic dimension due to
the small size of the sunspots. When individual local intrinsic dimension
estimates of the patches from these small sunspots are pooled with
the estimates from all other $\beta$ patches, the average intrinsic
dimension is more aligned with that of the other Mount Wilson types.
Additionally, ignoring the biggest outliers in the mean intrinsic
dimension (defined as having mean intrinsic dimension $>6.25$) gives
an average mean intrinsic dimension of 4.6 for the $\beta$ groups
which is more aligned with the other groups. 

The distribution of intrinsic dimension within the magnetic fragments
also differs by complexity based on Figures~\ref{fig:dimhist} and~\ref{fig:dimhist_pooled}.
The complex ARs have more patches and images with lower intrinsic
dimension than the simple sunspots which is consistent with Table~\ref{tab:dimresults}.

In summary, based on the estimated intrinsic dimension of the image
patches, relatively few parameters are required to accurately represent
the data. We have found that the distribution of local intrinsic dimension
varies based on the complexity of the sunspot group with the more
complex sunspots having higher (resp. lower) intrinsic dimension within
the sunspot (resp. magnetic fragments). Additionally, the standard
deviation of the intrinsic dimension is higher within the sunspot
in the simpler sunspots than the complex ARs. This is due to the presence
of small sunspots among the simpler ARs that tend to have less spatial
structure and thus a higher intrinsic dimension than typical sunspots.
We have also shown that linear methods should be sufficient to accurately
analyze the data.

\section{Spatial and Modal Correlations}

\label{sec:correlation}The results in the previous section indicate
that linear methods are likely sufficient to represent the spatial
and modal dependencies within a sunspot. We therefore analyze the linear correlation
over patches using partial correlation and canonical correlation analysis (CCA). 

The partial correlation is proportional to the inverse of the correlation matrix and analyzes the pixel-to-pixel correlation when the influence of all other pixels has been removed. It provides insight into how large a patch should be used to sufficiently capture the spatial and modal correlations in future analysis.

CCA on the other hand is determined by finding the most correlated linear combinations of pixels from each image, solved as a generalized eigenvalue problem, which is useful for determining the degree of mutual correlation between two modalities. If the two modalities are independent, there is no benefit in processing them together, while if the two modalities are strongly dependent, processing only one of the modalities is sufficient since the other modality would not contain any additional information.

\subsection{Partial Correlation: Methodology}

\label{sub:partialcorr}The partial correlation measures the correlation
between two random variables while conditioning on the remaining random
variables. The intuition behind partial correlation  can be best explained with the linear regression concept. Suppose you want to compute the partial correlation between two variables $X_1$ and $X_2$ given a set of variables $\mathcal{X}$. First, compute the linear regression using variables in $\mathcal{X}$ to explain $X_1$ and obtain the associated residuals $r_{X_1}$. Proceed similarly for $X_2$ and get residuals $r_{X_2}$. The partial correlation between $X_1$ and $X_2$ is then equal to the (usual) correlation between $r_{X_1}$ and $r_{X_2}$, for which the effect of variables $\mathcal{X}$ have been removed.  

In our context, let $\mathbf{x}$ be a patch from the continuum image, and $\mathbf{|y}|$
be the magnitude (entry-wise absolute value) of the corresponding
patch from the magnetogram. The partial correlation matrix $\mathbf{P}=\left(\begin{array}{cc}
\mathbf{P}_{\mathbf{xx}} & \mathbf{P}_{\mathbf{x|y|}}\\
\mathbf{P}_{\mathbf{|y|x}} & \mathbf{P}_{\mathbf{|y||y|}}
\end{array}\right)$ and its off-diagonal elements can be derived from the inverse correlation matrix (see Appendix~\ref{sub:partcorr}).
We use the magnitude of the magnetogram data since both positive and
negative polarities affect the continuum image in similar ways.

\subsection{Partial Correlation: Results}
\begin{figure}
\centering

\includegraphics[width=0.5\textwidth]{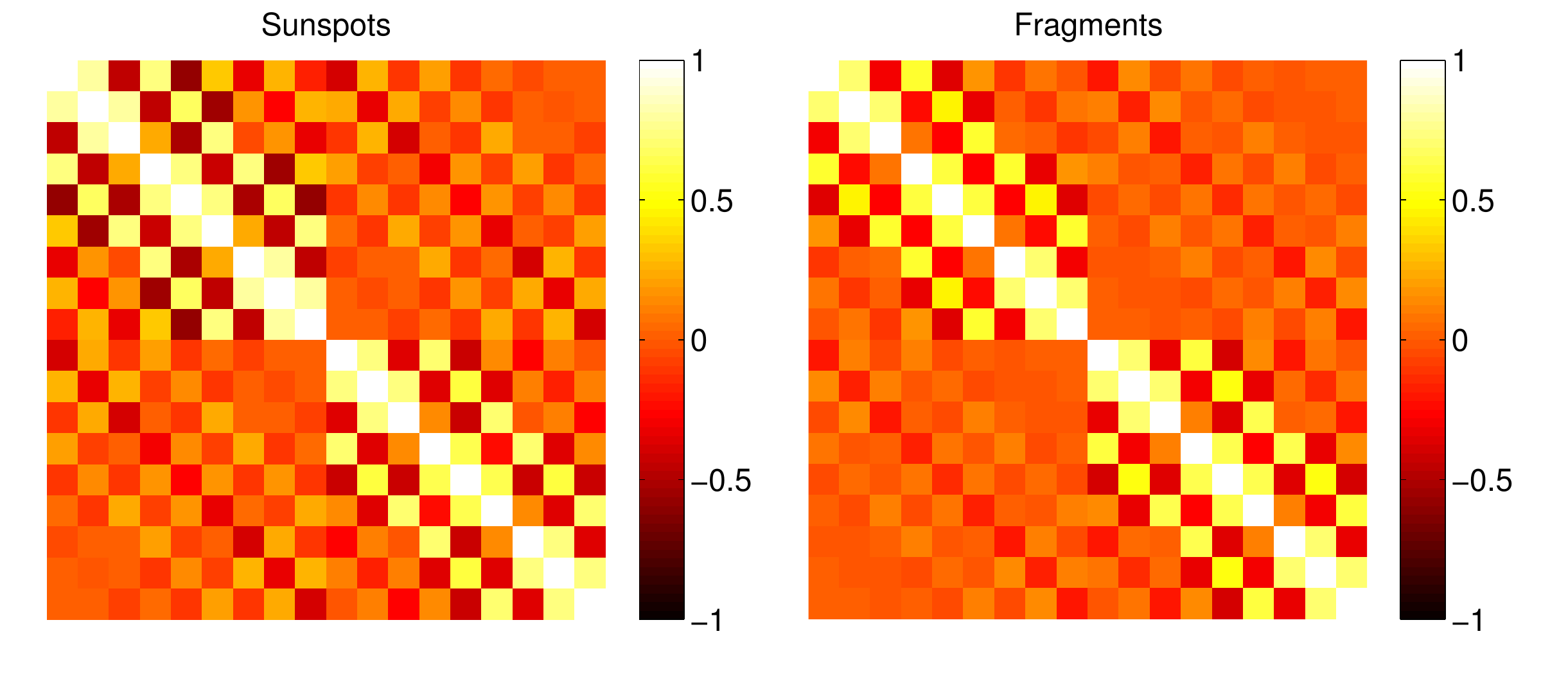}\includegraphics[width=0.5\textwidth]{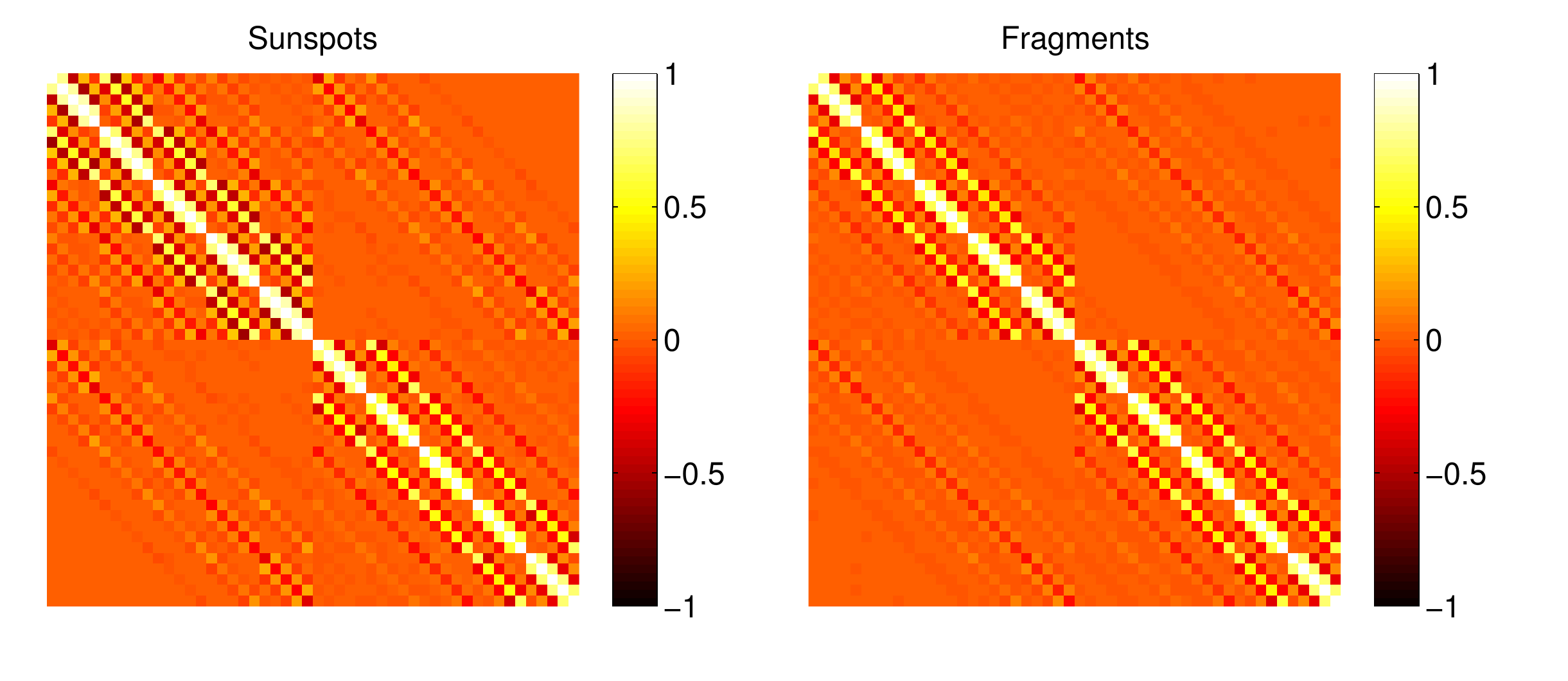}

\caption{Estimated partial correlation matrices of patch data from within the
sunspots and the magnetic fragments using $3\times3$ (left) and $5\times5$
(right) patches. The theoretical thresholds~\citep{hero2011corr}
for significance to attain a $0.05$ false alarm rate are $0.0070$
and $0.0014$ for within the sunspots and magnetic fragments, respectively
when using a $3\times3$ patch. For the $5\times5$ patch, the thresholds
are $0.0080$\textbf{ }and $0.0016$, respectively. Statistically
insignificant values are set to zero.\label{fig:partial}}
\end{figure}

Figure~\ref{fig:partial} gives the estimated partial correlation
matrices when using $3\times3$ and $5\times5$ patches. The patches
are extracted from all of the active regions and divided using the
STARA and SMART masks into sunspots and magnetic fragments as before. The partial correlation of $3\times3$ patches
is quite strong within both modalities. Based on a false alarm rate
of $0.05$, the theoretical thresholds for significance for the partial
correlation~\citep{hero2011corr} of the $3\times3$ patches are
approximately $0.0070$ and $0.0014$ for within the sunspots and
magnetic fragments, respectively. For the $5\times5$ patches, the thresholds are $0.0080$ and $0.0016$, respectively. Given these thresholds, the partial
correlation is statistically significant for nearly all values within
the modalities ($\mathbf{P}_{\mathbf{xx}}$ and $\mathbf{P}_{\mathbf{|y||y|}}$) using the $3\times3$ patches.

The cross-partial correlation when using the signed magnetic field
(i.e. $\mathbf{P}_{\mathbf{xy}}$ and $\mathbf{P}_{\mathbf{yx}}$) is very near zero
in both regions (not shown). However, when we take the absolute value
of the magnetogram patches, then the magnitude of the cross-partial
correlation ($\mathbf{P}_{\mathbf{x|y}|}$ and $\mathbf{P}_{|\mathbf{y|x}}$) is much
higher in both regions suggesting that the correlation between the
modalities is significant. The partial correlation is also stronger
in magnitude in all cases within the sunspots than within the magnetic
fragments. 

The partial correlation matrices are very structured. In both
sunspots and magnetic fragments, the pentadiagonal-like structure
within the modalities suggests that the image is generally stationary
with approximately a third order nearest neighbor Markov structure in the pixels.
Such structure is clearly seen in the matrices for $5\times5$ patches.
The cross-partial correlation also has a pentadiagonal-like structure
although the correlation is not as strong as within the modalities. 

\begin{figure}
\centering

\includegraphics[width=0.75\textwidth]{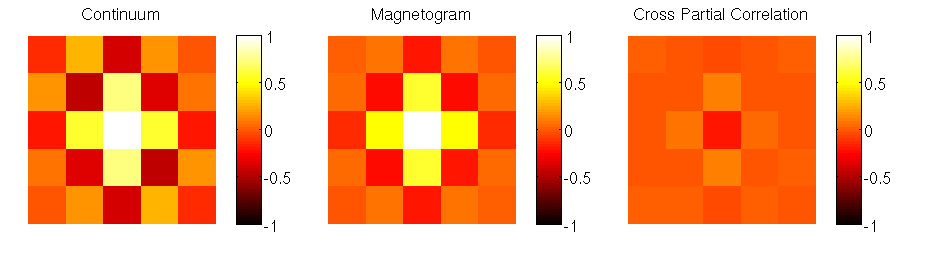}

\caption{Partial correlation patches extracted from the columns in the sunspot
partial correlation matrix corresponding to the center pixels. The partial correlation is stronger within the continuum. \label{fig:partial_patches}}

\end{figure}

To better see the spatial correlations, in Figure~\ref{fig:partial_patches}
we plot the partial correlation patches taken from the columns of
the sunspot partial correlation matrix corresponding to the center
pixels when using $5\times5$ patches. Figure~\ref{fig:partial_patches} shows clearly the greater partial correlation within the continuum. It also highlights that correlation is slightly higher in magnitude in the vertical direction than the horizontal direction.  Nearly all sunspots in this study are located within $(-30^\circ, + 30^\circ)$ from both the central meridian and the equator, and so projection effect are unlikely to cause this difference. The difference in correlation may be a feature of the sunspots themselves, but this may be difficult to determine since the difference in partial correlation is small.

\begin{figure}
\centering

\includegraphics[width=0.9\textwidth]{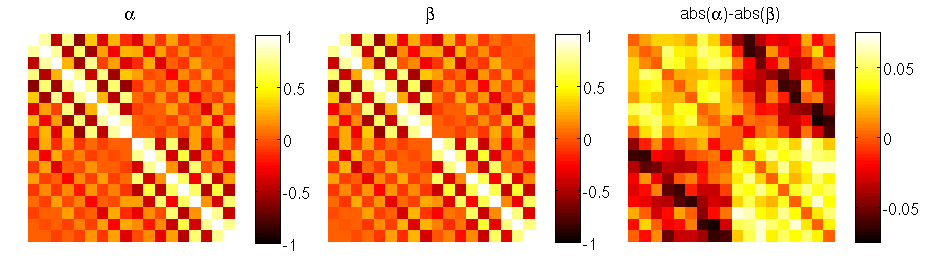}

\caption{Partial correlation matrices within the sunspots using the data from
$\alpha$ (left) and $\beta$ ARs (center). Statistically insignificant
values are again set to zero. (Right) difference between the absolute
value of the $\alpha$ and $\beta$ matrices. The $\alpha$ sunspots
are more (resp. less) strongly correlated within (resp. between) the
modalities than the $\beta$ groups.\label{fig:Partmwilson} }

\end{figure}

Some slight differences exist in the partial
correlation matrices restricted to certain Mount Wilson classes. As an example, Figure~\ref{fig:Partmwilson} contains the partial correlation
matrices within the sunspots after restricting the data to $\alpha$
and $\beta$ groups as well as the difference between the absolute
value of the two matrices. The $\alpha$ partial correlation matrix
is higher in magnitude within the modalities than the $\beta$ matrix
but lower between the modalities. Within modalities, the strongest
differences (a maximum of 0.056 and 0.067 within the continuum and
magnetogram, respectively) are in the entries that correspond to pixels
that are farther away from each other. In contrast, within the cross-partial
correlation, the strongest differences (a maximum of 0.072) between
the two AR types are in the entries that correspond to pixels that
are close to each other. A similar pattern holds when comparing the
$\alpha$ matrix to the matrices of the more complex groups.

Overall, the partial correlation matrices indicate that no larger
than a $5\times5$ patch is necessary to capture the local spatial
dependencies. A $5\times5$ patch of pixels corresponds roughly to the size of a mesogranule \citep{rast2003scales,rieutord2000mesogranulation}. This suggests that within the magnetic fragments, it is likely that the granules and mesogranules within the photosphere contribute to the local spatial dependencies. Within the sunspots, a $5\times 5$ patch corresponds to the size of the characteristic length of the largest penumbral filaments \citep{tiwari2013structure} which suggests that on average the local spatial dependencies are minimal beyond this scale. This analysis, however, does not rule out long-range spatial dependencies, which are more difficult to assess due to the large dimensionality. Future work will focus on this.

In the remainder of our analysis, we choose
a $3\times3$ patch for the reasons mentioned in Section~\ref{sec:data}: to ensure that we capture the features of small sunspots and to limit the effects high dimension on the accuracy of the analysis. Given these
concerns, we see that $3\times3$ patches capture most of the spatial
correlation. This is evident from Figure~\ref{fig:partial} where the partial correlation between pixels on opposite corners
of a $3\times3$ patch is near zero and other pixels that are
similarly far away from each other have low partial correlation. Thus
a $3\times3$ patch strikes a good balance between scale, extrinsic
dimension, and capturing the spatial correlation.

\subsection{Canonical Correlation Analysis: Methodology}

To further investigate the correlation between the modalities, we
use canonical correlation analysis (CCA) on the continuum patch $\mathbf{x}$
and the magnitude of the magnetogram patch $\mathbf{y}$. CCA finds
patterns and correlations between two multivariate data sets \citep{muller82,nimon:cca}
and was used previously in the context of space weather e.g. for the
combined analysis of solar wind and geomagnetic index data sets \citep{2014JGRA..119.5364B}. 

In our application, CCA provides linear combinations of continuum patches $\mathbf{x}$ that are most correlated with linear combinations of magnetogram patches $\mathbf{y}$. In other words, all correlations between the continuum and magnetogram patches are channeled through the canonical variables. Formally, CCA  finds vectors $\mathbf{a}_{i}$ and $\mathbf{b}_{i}$ for $i=1,\dots,m^2$ such that
the correlation $\rho_{i}=\text{corr}(\mathbf{a}_{i}^{T}\mathbf{x},\mathbf{b}_{i}^{T}|\mathbf{y}|)$
is maximized and the pair of random variables $u_{i}=\mathbf{a}_{i}^{T}\mathbf{x}$
and $v_{i}=\mathbf{b}_{i}^{T}|\mathbf{y}|$ are uncorrelated with all other
pairs $u_{j}$ and $v_{j}$, $j\neq i$. The variables $u_{i}$ and
$v_{i}$ are called the $i$th pair of canonical variables while the
vectors $\mathbf{a}_{i}$ and $\mathbf{b}_{i}$ are the canonical vectors. The solution
$\mathbf{a}_{i}$ is the $i$th eigenvector of the matrix $\Sigma_{\mathbf{xx}}^{-1}\Sigma_{\mathbf{x|y}|}\Sigma_{|\mathbf{y||y|}}^{-1}\Sigma_{|\mathbf{y|x}}$ which are taken from the covariance matrix.
The vector $\mathbf{b}_{i}$ is found similarly~\citep{hardle2007applied}.

\subsection{Canonical Correlation Analysis: Results}
\begin{figure}
\centering

\includegraphics[width=0.33\textwidth]{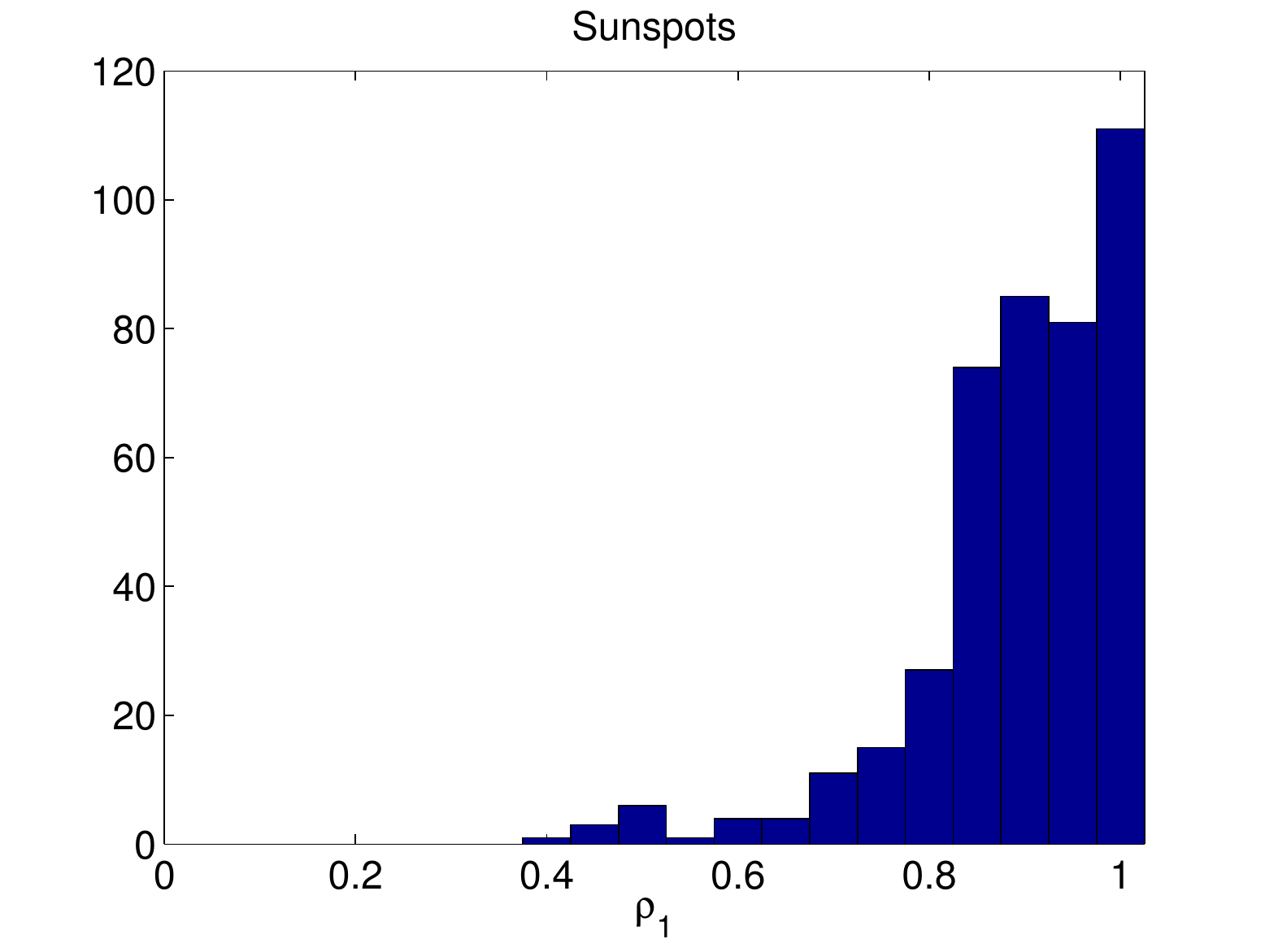}\includegraphics[width=0.33\textwidth]{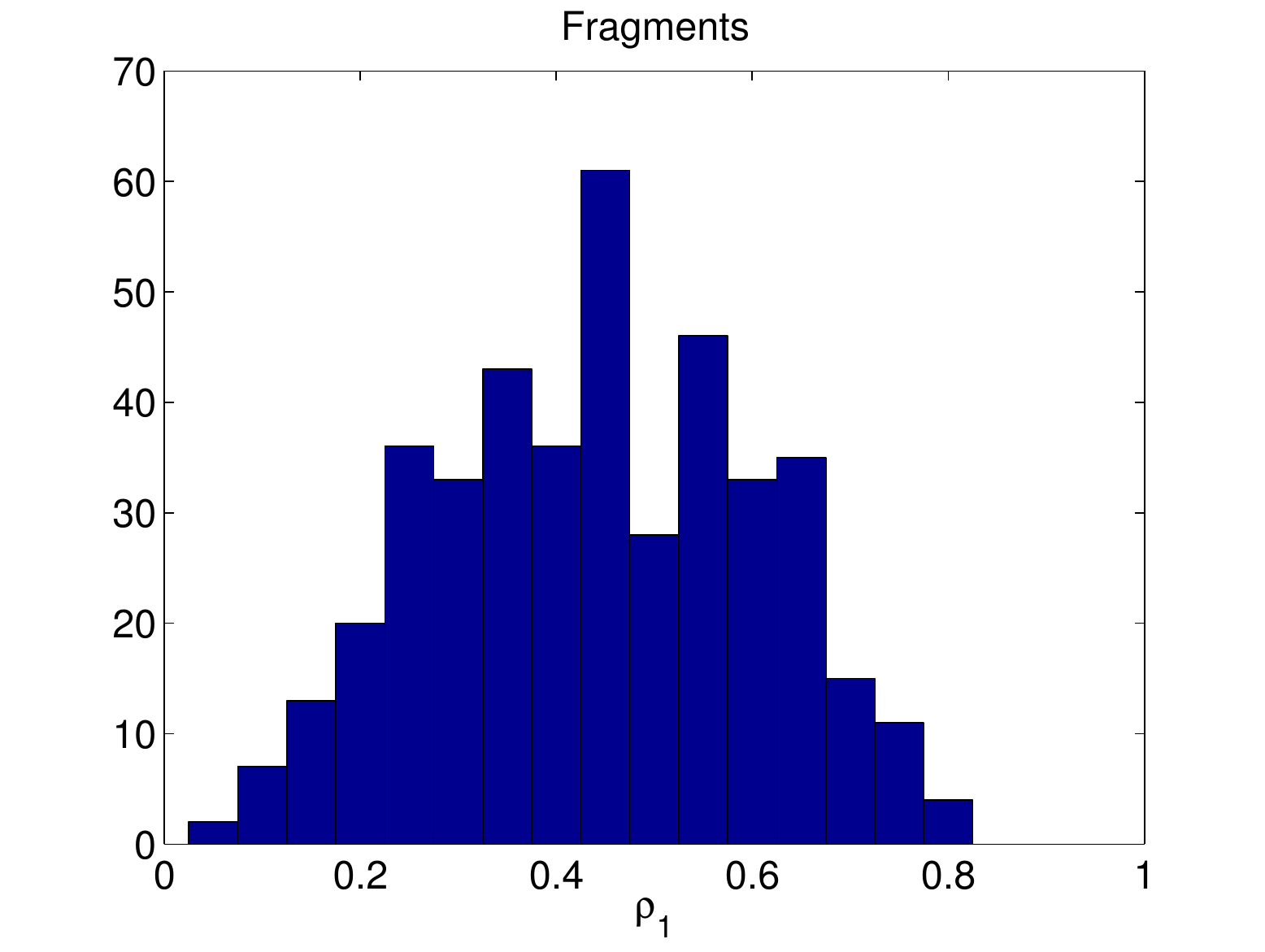}\includegraphics[width=0.33\textwidth]{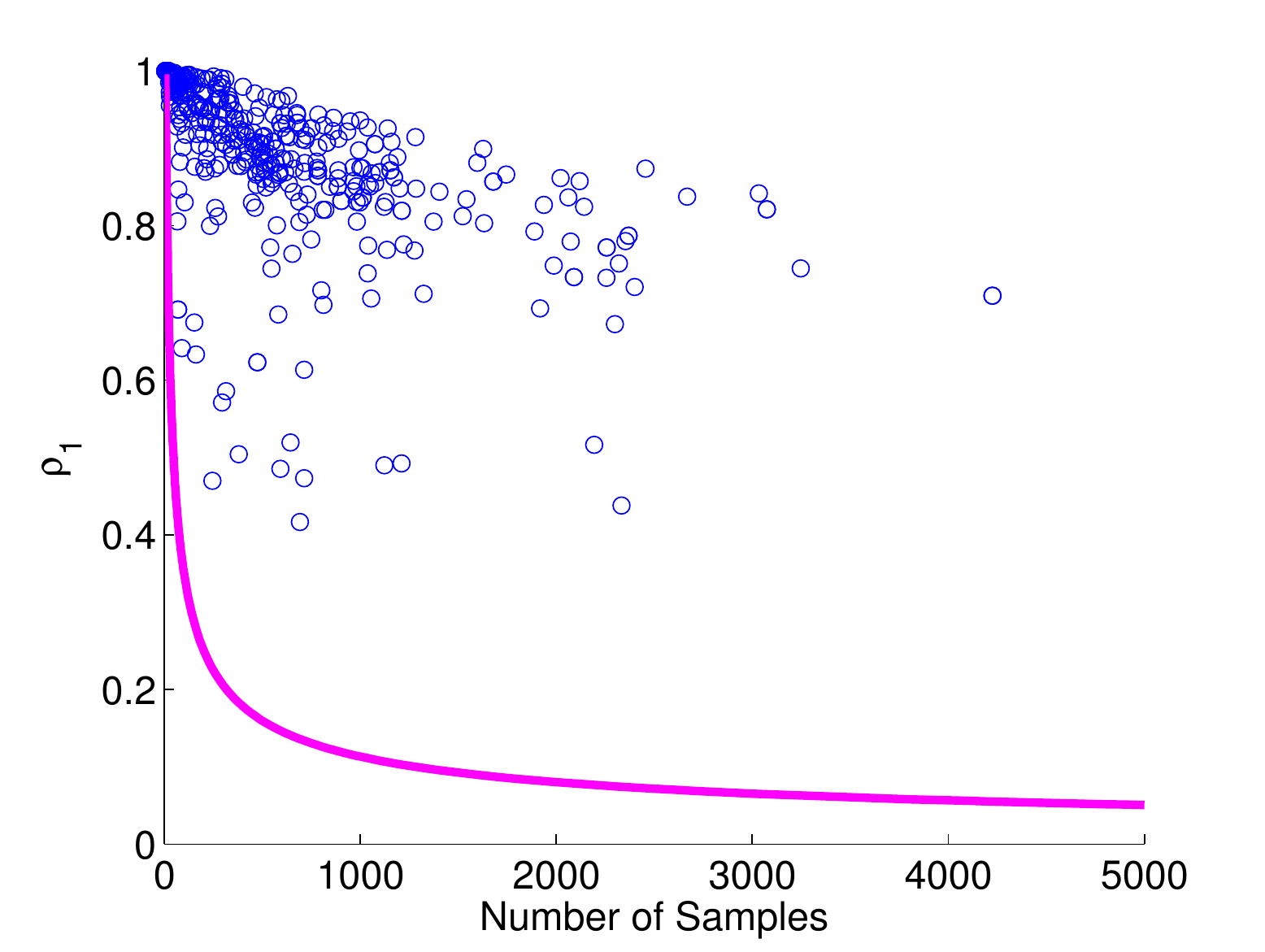}

\caption{Histograms of estimated $\rho_{1}$ using CCA for within the sunspots
(left) and the magnetic fragments (center) using $3\times3$ patches.
Right: Scatter plot of $\rho_{1}$ values and the number of samples
available for within the sunspots. All points are above the magenta
line which gives the threshold for statistical significance at a false alarm rate of 0.05 \citep{hero2011corr}.\label{fig:CCA1}}

\end{figure}

Here we focus on $3\times3$ patches and apply CCA to all 424 image
pairs using the magnitude of the magnetogram patches. Figure~\ref{fig:CCA1}
(left and center) shows histograms of the estimated values of $\rho_{1}$.
Within the sunspots, there are many groups with a near perfect correlation
between the modalities and none of the groups have an estimated value
below $0.41$. The right plot in Figure~\ref{fig:CCA1} plots the estimated
values of $\rho_{1}$ vs. the number of samples used within the sunspots.
Based on this plot, there are many ARs with high correlation
and few patch samples suggesting that the correlation may be spurious. However,
all of the estimated values are statistically significant as defined
by the threshold given by \cite{hero2011corr} using a false alarm rate of 0.05 (shown as the magenta
line in Figure~\ref{fig:CCA1}). 

The histogram of $\rho_{1}$ within the magnetic fragments (Fig.~\ref{fig:CCA1},
center) is quite different from the sunspot histogram (Fig.~\ref{fig:CCA1}, left). $\rho_{1}$
is generally lower within the magnetic fragments than within the sunspots
which is consistent with the results in Figure~\ref{fig:partial}.
All of the $\rho_{1}$ values are statistically significant.

\begin{figure}
\centering

\includegraphics[width=0.5\textwidth]{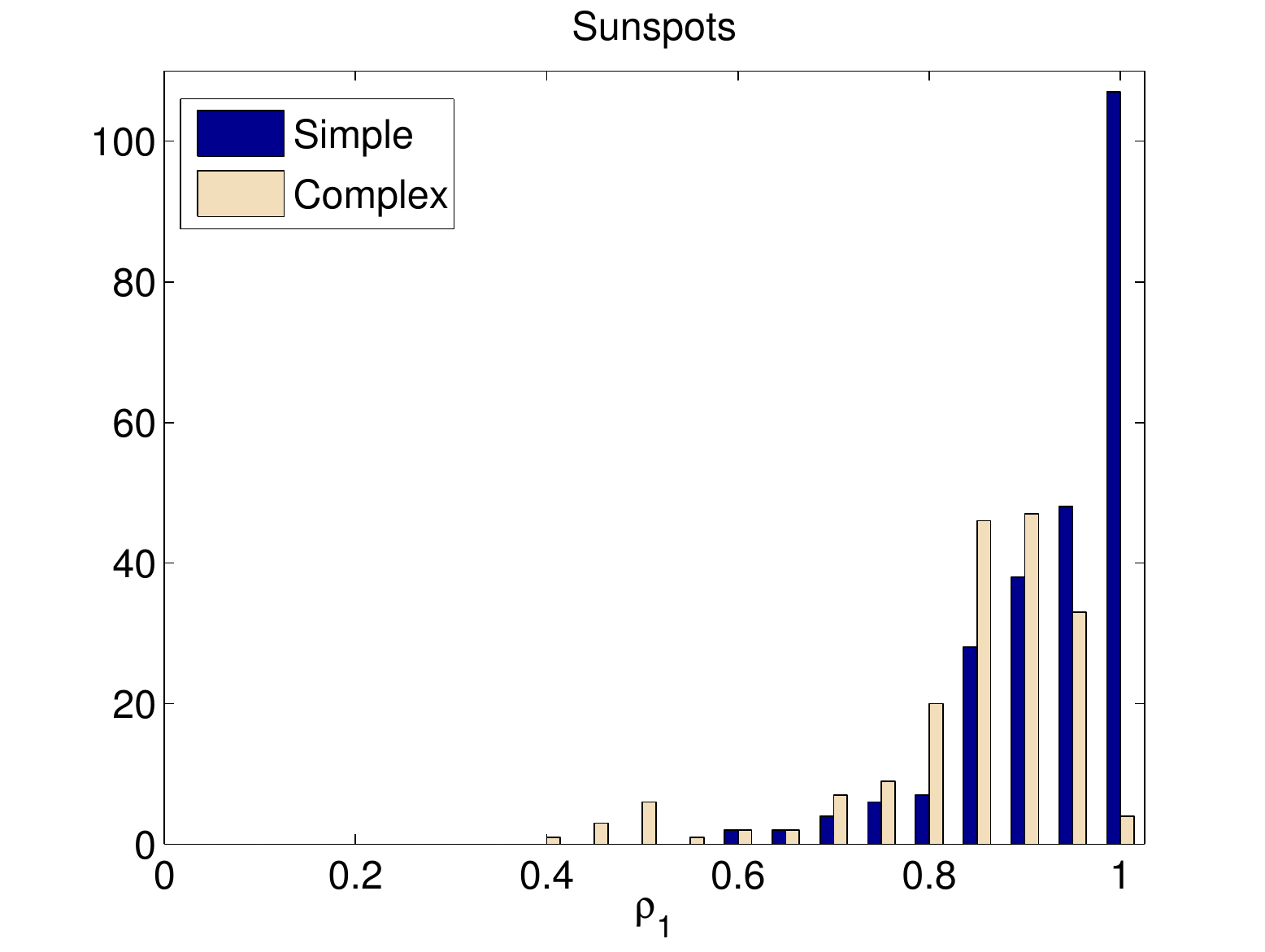}\includegraphics[width=0.5\textwidth]{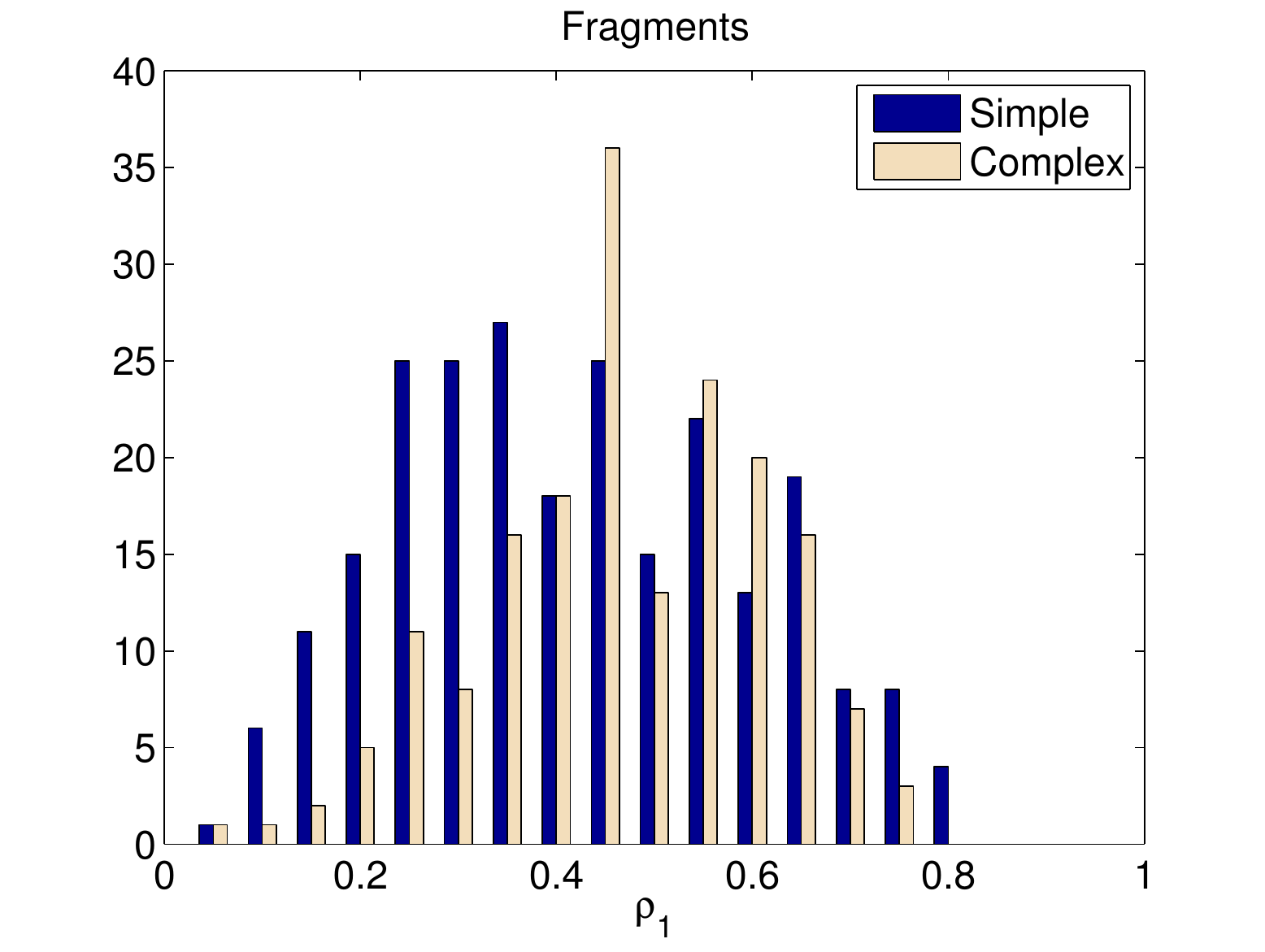}

\caption{$\rho_{1}$ histograms of complex ($\beta\gamma$ and $\beta\gamma\delta$)
and simple ($\alpha$ and $\beta$) regions within the sunspots (left)
and the magnetic fragments (right). The simple ARs are generally more
correlated within the sunspots but less correlated within the magnetic
fragments. The difference between the sunspot distributions, as measured
by the Hellinger distance, is statistically significant.\label{fig:histCCA}}

\end{figure}

The distributions of $\rho_{1}$ differ slightly when comparing simple
sunspot groups ($\alpha$ and $\beta$) with complex groups ($\beta\gamma$
and $\beta\gamma\delta$). Figure~\ref{fig:histCCA} shows that complex
groups generally have lower correlation between the modalities within
the sunspots than the simpler groups. The estimated Hellinger distance
(see Appendix~\ref{sub:hellinger}) between the distributions using
the divergence estimator in~\cite{moon2014isit,moon2014nips} is
0.22. Based on the central limit theorem of the estimator \citep{moon2014nips},
this value is statistically significant\textbf{ }with a $p$-value
of $1.6\times10^{-12}$. At least some of this difference is likely
due to the smaller size of the simpler groups (and thus smaller sample
size). However, it is unlikely to fully explain the difference given
that there are many simple sunspot groups with high correlation and
sufficient sample size.

Within the magnetic fragments, there are many more simple regions
than complex regions with $\rho_{1}<0.4$ (see the histogram in Figure~\ref{fig:histCCA},
right). This could be related to the same phenomena that causes the
intrinsic dimension to be higher within the magnetic fragments of
simple sunspot groups observed in Section~\ref{sub:dimresults}.
However, the estimated Hellinger distance between these distributions
is 0.016. Using the same statistical test, this estimate is not statistically
significant with a $p$-value of $0.31$. Thus the distributions are
not statistically different from each other.

\begin{figure}
\centering

\includegraphics[width=0.35\textwidth]{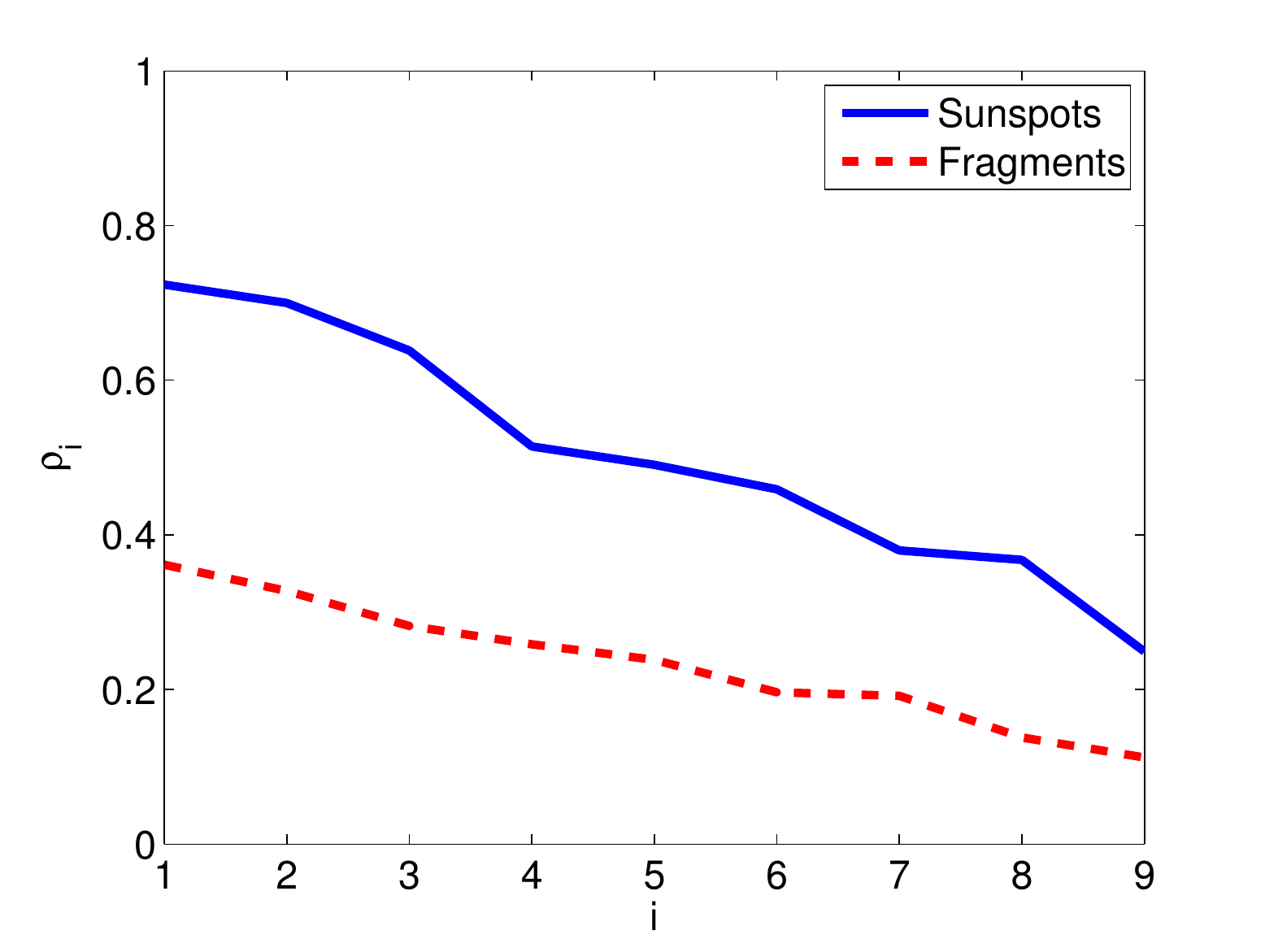}\includegraphics[width=0.65\textwidth]{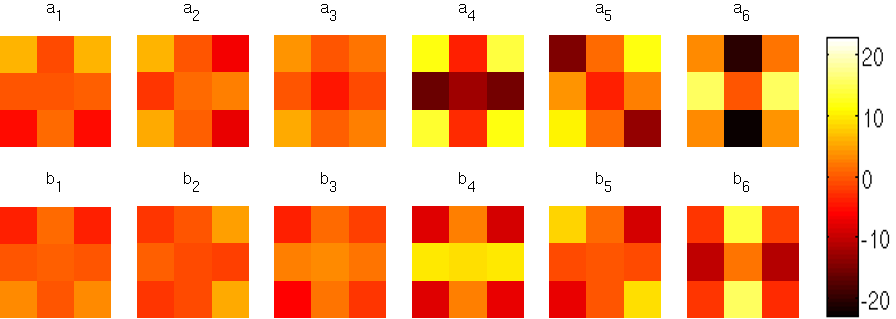}

\caption{\label{fig:patchCCA}(Left) Plot of the estimated $\rho_{i}$ using
CCA on the entire data set for $i=1,\dots,9$. All values are statistically
significant~\citep{hero2011corr}.\label{fig:allCCA} (Right) Canonical
patches $\mathbf{a}_{i}$ (top) and $\mathbf{b}_{i}$ (bottom) for $i=1,\dots,6$ when
using the entire data set from within the sunspots. The $\mathbf{b}_{i}$s
are approximately equal to the negative of the $\mathbf{a}_{i}$s.}
\end{figure}

To analyze the spatial patterns that produce the highest correlation
between modalities, we apply CCA to the entire data set. Figure~\ref{fig:allCCA}
plots $\rho_{i}$ for $i=1,\dots,9$ for within the sunspots and within
the magnetic fragments. The $\rho_{i}$ are all statistically significant.
Notice that the $\rho_{i}$ are higher within the sunspots than the
magnetic fragments which is consistent with the results in Figures~\ref{fig:partial}
and~\ref{fig:CCA1}. 

Figure~\ref{fig:patchCCA} shows the canonical patches $\mathbf{a}_{i}$ and
$\mathbf{b}_{i}$ for $i=1,\dots,6$ when using all the data from within the
sunspots. These are the spatial patterns within the two modalities
that are most correlated with each other. The canonical patches have
a ``saddle-like'' appearance where the gradient is positive in some
directions and negative in others. For example, in $\mathbf{a}_{4}$, the pixels
to the left and right of the center are very negative but the pixels
in the corners are all very positive. Note that these vectors correspond to centered values with respect to the mean patches.

Comparing the $\mathbf{a}_{i}$s to the $\mathbf{b}_{i}$s shows that the $\mathbf{b}_{i}$s
are approximately equal to the negative of the $\mathbf{a}_{i}$s. This makes
sense as sunspots within the continuum images correspond to a decrease
in value relative to the background while ARs within the magnitude
of the magnetogram images correspond to an increase in value relative
to the background.

\begin{figure}
\centering

\includegraphics[width=0.35\textwidth]{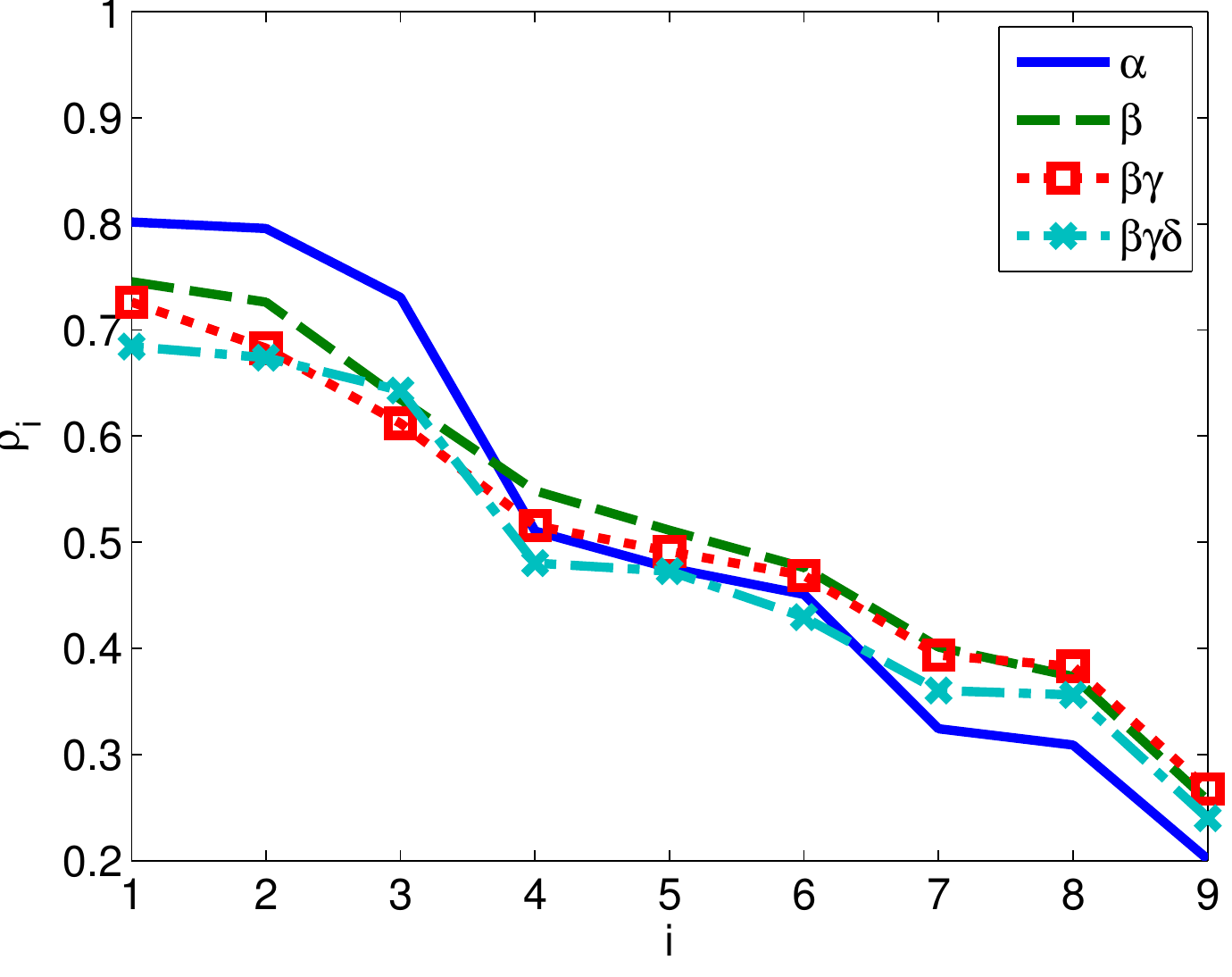}\hspace{1cm}\includegraphics[width=0.45\textwidth]{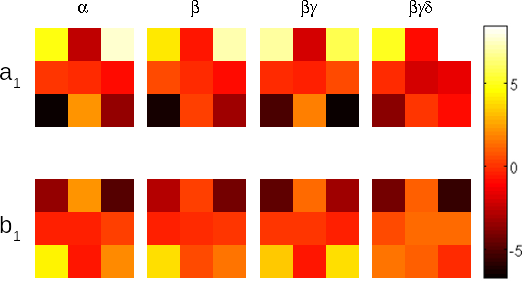}

\caption{(Left) Plot of the estimated $\rho_{i}$ using CCA on data segregated
by Mount Wilson classes for $i=1,\dots,9$ within the sunspots $\alpha$
groups start out with the highest correlation. (Right) Canonical patches
$\mathbf{a}_{1}$ (top) and $\mathbf{b}_{1}$ (bottom) for the Mount Wilson classes
within the sunspots. Again, the $\mathbf{b}_{1}$s are approximately equal
to the negative of the $\mathbf{a}_{1}$s as in Figure~\ref{fig:patchCCA} but
the patches differ slightly from class to class.\label{fig:CCAMtwilson}}

\end{figure}

We also performed CCA separately on the data from the Mount Wilson
classes. Figure~\ref{fig:CCAMtwilson} plots the $\rho_{i}$ values
for each class and the first canonical patches $\mathbf{a}_{1}$ and $\mathbf{b}_{1}$.
For $\rho_{1}$ and $\rho_{2}$, the values for each class decrease
in order of complexity ($\alpha,$ $\beta$, $\beta\gamma,$ $\beta\gamma\delta$).
This is consistent with our comparison of the partial correlation
matrices in Figure~\ref{fig:Partmwilson} where the partial correlation
was generally higher (in magnitude) for the $\alpha$ groups than
the others. This is also consistent with the intrinsic dimension analysis
in Section~\ref{sec:dimension} where the intrinsic dimension generally
increases with complexity. This is because if the correlation between
and within modalities is higher, then fewer parameters are required
to accurately describe the data which results in a lower intrinsic
dimension.

The canonical patches $\mathbf{a}_{1}$ and $\mathbf{b}_{1}$ have similar patterns
across the different classes although the patches for the $\beta\gamma$
class are flipped compared to the others. The magnitude of the values
in the $\beta\gamma\delta$ patches are also smaller than the those
of the other patches.

Overall, the results of this section suggest that the two modalities
are correlated in both the sunspots and the magnetic fragments and
are therefore not independent. The correlation is stronger within
the sunspots compared to the magnetic fragments and stronger within
the sunspots in simple ARs compared to complex ARs. However, the correlation
is not perfect and so there may be an advantage to including both
modalities in the classification of sunspots and flare prediction.

\section{Conclusion}

Existing AR categorical classification systems such as the Mount Wilson
and McIntosh schemes describe geometrical arrangements of the magnetic field at the \emph{largest} length scale. In this work, we have focused on the properties
of the ARs at \emph{fine} length scale. We showed that when we analyze the global statistics or
attributes of these local properties, we find differences between
the simple and complex ARs as defined using the large scale characteristics.
So by this approach, we are analyzing both the large and fine scale properties
of the images. Such results might be due to the multi-scale properties of the magnetic fields, as evidenced previously in~\cite{Ireland2008mra}.

The local intrinsic dimension based on the $k$-NN approach combines
both continuum and magnetogram observations and provides some measure
of local regularity for those images. Further differences between the Mount Wilson classes may be found by comparing the histograms or distributions of local intrinsic dimension of each individual AR instead of only comparing the means or pooled estimates as we did in this paper. There are several options to perform such comparisons. Each histogram could be treated as a vector, or we could consider the underlying probability density function within the framework of functional analysis. Supervised (using Mount Wilson classes) or unsupervised classification could be performed. Another option would be to view the set of histograms belonging to a specific class as samples from a distribution of vectors (or a distribution of probability density functions). Different classes could then be compared using divergence measures such as the Hellinger distance described in Appendix~\ref{sub:hellinger}.

This work also highlighted specific behaviors of the core of active
regions (that corresponds to the sunspot masks in continuum) and magnetic
fragments (the surrounding part of AR), as well as the difference
of these two regions as a function of the Mount Wilson classification.
We found that within the sunspots, the spatial and modal correlations are stronger than within the magnetic fragments. Additionally, simpler ARs were found to have higher correlation between the modalities within the sunspots than the complex ARs. 

This study paves the way for further analysis based on dictionary
learning. Knowledge of the intrinsic dimension allows us to choose
the dictionary size. Moreover the results of Section~\ref{sec:dimension}
showed that linear dictionary learning methods are sufficient. The
spatial and modal correlation analysis in Section~\ref{sec:correlation}
justifies a choice of a patch size of $3\times3$ and confirms that
both modalities (continuum and magnetogram) should be used in dictionary
learning. 

\begin{acknowledgements}
This work was partially supported by the US National Science Foundation (NSF) under grant CCF-1217880 and a NSF Graduate Research Fellowship to KM under Grant No. F031543. VD acknowledges support from the Belgian Federal Science Policy Office through the ESA-PRODEX program, grant No. 4000103240, while RDV acknowledges support from the BRAIN.be program of the Belgian Federal Science Policy Office, contract No. BR/121/PI/PREDISOL. The editor thanks two anonymous referees for their assistance in evaluating this paper.
\end{acknowledgements}

\begin{appendix}

\section{Method Details}


\subsection{Intrinsic Dimension Estimation of Manifolds}

\label{sub:knn}Consider data that are described in an extrinsic Euclidean
space of $d$ dimensions. However, suppose the data actually lie on
a lower dimensional manifold $\mathcal{M}$. Thus the intrinsic dimension
$m$ of the data corresponds to the dimension of $\mathcal{M}$. For
example, data may be given to us in a 3 dimensional space but lie
on the surface of a sphere. Thus the intrinsic dimension of the data
would be 2. 

In some cases, data points from the same data set may lie on different
manifolds. For example, part of the data with an extrinsic dimension
of 3 could lie on the surface of a sphere ($m=2$) while another part
may lie on a circle ($m=1$). We then say that data points from these
different manifolds have a different \emph{local }intrinsic dimension.
The local intrinsic dimension gives some measure of the local complexity
of the image. Additionally, the local intrinsic dimension is useful
for dictionary learning because we can use it to determine whether
different-sized dictionaries should be used for different regions,
e.g. within the sunspots and outside of the sunspots. 

We now describe the $k$-NN estimator of intrinsic dimension in more
detail. For a set of independently identically distributed random
vectors $\mathbf{Z}_{n}=\{\mathbf{z}_{1},\dots,\mathbf{z}_{n}\}$ with values in a compact
subset of $\mathbb{R}^{d}$, the $k$-nearest neighbors of $\mathbf{z}_{i}$
in $\mathbf{Z}_{n}$ are the $k$ points in $\mathbf{Z}_{n}\backslash\{\mathbf{z}_{i}\}$
closest to $\mathbf{z}_{i}$ as measured by the Euclidean distance $||\cdot||$.
The $k$-NN graph is then formed by assigning edges between a point
in $\mathbf{Z}_{n}$ and its $k$-nearest neighbors. The intrinsic
dimension is related to the total edge length of the $k$-NN graph
and can be estimated based on this relationship. The $k$-NN graph is then formed
by assigning edges between a point in $\mathbf{Z}_{n}$ and its $k$-nearest
neighbors and has total edge length defined as 
\[
L_{\gamma,k}(\mathbf{Z}_{n})=\sum_{i=1}^{n}\sum_{\mathbf{z}\in\mathcal{N}_{k,i}}||\mathbf{z}-\mathbf{z}_{i}||^{\gamma},
\]
where $\gamma>0$ is a power weighting constant and $\mathcal{N}_{k,i}$
is the set of $k$ nearest neighbors of $\mathbf{z}_{i}.$ It has been shown
that for large $n$,
\[
L_{\gamma,k}(\mathbf{Z}_{n})=n^{\alpha(m)}c+\epsilon_{n},
\]
where $\alpha=(m-\gamma)/m$, $c$ is a constant with respect to $\alpha(m)$,
and $\epsilon_{n}$ is an error term that decreases to zero a.s. as
$n\rightarrow\infty$ \citep{costa2006determining}. A global intrinsic
dimension estimate $\hat{m}$ is found based on this relationship
using non-linear least squares over different values of $n$ \citep{carter2010local}.

A local estimate of intrinsic dimension at a point $\mathbf{z}_{i}$ can be
found by running the algorithm over a smaller neighborhood about $\mathbf{z}_{i}.$
The variance of this local estimate is then reduced by smoothing via
majority voting in a neighborhood of $\mathbf{z}_{i}$ \citep{carter2010local}.

\subsection{Partial Correlation}

\label{sub:partcorr}Let $\mathbf{z}$ be a random vector with size
$m$. Let $\Sigma$ be the covariance matrix of $\mathbf{z}$, that
is $\Sigma_{ij}=\text{Cov}(z_{i},z_{j})$, and let $\mathbf{K}=\Sigma^{-1}$
be the inverse of the covariance matrix, also called the precision
matrix.

The partial correlation between $z_{i}$ and $z_{j}$ given all the
other variables $\mathbf{z}\backslash\{z_{i},z_{j}\}$ measure the
degree of correlation between these two variables after removing the
effect of the remaining ones. Let $\mathbf{P}_{ij}$ denote the partial correlation
between $z_{i}$ and $z_{j}$. It has been shown \citep{lauritzen1996graphical}
that $\mathbf{P}_{ij}$ can be related to the elements of the precision matrix
$\mathbf{K}$ as follows: 
\[
\begin{array}{cc}
\mathbf{P}_{ij}=-\frac{K_{ij}}{\sqrt{K_{ii}K_{jj}}}, & i\neq j.\end{array}
\]

\subsection{Estimating the Hellinger Distance}

\label{sub:hellinger}Information
divergences are a class of functionals that measure the difference
between two probability distributions. The most popular divergence
measure is the Kullback-Leibler divergence \citep{kullback1951div}.
The Hellinger distance is another divergence measure and is defined
as 
\[
H(f,g)=\sqrt{1-\int\sqrt{f(x)g(x)}dx},
\]
where $f$ and $g$ are the two probability densities being compared.
The Hellinger distance is a metric which is not true of divergences
in general. we use the nonparametric divergence estimator derived
in \cite{moon2014isit,moon2014nips}. In \cite{moon2014isit}, it
was shown that this estimator converges to the true divergence with
mean squared error convergence rate $O(1/T)$ where $T$ is the number
of samples from each probability distribution. In \cite{moon2014nips},
it was shown that the distribution of the normalized version of this
estimator converges to the standard normal distribution. We can use
this fact combined with a bootstrap estimate~\citep{Efron1994} of the variance of the
estimator to test the hypothesis that the divergence is zero (and
hence the distributions are equal). 

\end{appendix}

\bibliographystyle{swsc}
\bibliography{SWSC_Moon}

\end{document}